\documentclass[aps,nofootinbib,preprint,superscriptaddress]{revtex4-2}%
\usepackage{hyperref}
\usepackage{amsmath}
\usepackage{amsfonts}
\usepackage{amssymb}
\usepackage{graphicx}
\usepackage{color}

\allowdisplaybreaks[2]

\begin{document}

\title{The stringy scaling loop expansion and stringy scaling violation}

\author{Sheng-Hong Lai}
\email{xgcj944137@gmail.com}
\affiliation{Department of Electrophysics, National Yang Ming Chiao Tung University, Hsinchu, ROC}
\author{Jen-Chi Lee}
\email{jcclee@nycu.edu.tw}
\affiliation{Department of Electrophysics, National Yang Ming Chiao Tung University, Hsinchu, ROC}
\author{Yi Yang}
\email{yiyang@nycu.edu.tw}
\affiliation{Department of Electrophysics, National Yang Ming Chiao Tung University, Hsinchu, ROC}
\affiliation{Center for Theoretical and Computational Physics, National Yang Ming Chiao Tung University, Hsinchu, ROC}

\date{\today}

\begin{abstract}
We propose a systematic approximation scheme to calculate general
string-tree level $n$-point hard string scattering amplitudes ($HSSA$) of
open bosonic string theory. This stringy scaling loop expansion contains
finite number of vacuum diagram terms at each loop order of scattering
energy due to a vacuum diagram contraint and a topological graph constraint.
In addition, we calculate coefficient and give the vacuum diagram
representation and its Feynman rules for each term in the expansion of the $%
HSSA$. As an application to extending our previous calculation of $n$-point
leading order stringy scaling behavior of $HSSA$, we explicitly calculate
some examples of $4$-point next to leading order stringy scaling violation
terms.
\end{abstract}

\maketitle

\newpage

\setcounter{equation}{0} \renewcommand{\theequation}{\arabic{section}.\arabic{equation}}

\section{Introduction}

In contrast to the finite number of coupling vertices in field theory, there
are infinite $n$-point coupling vertices with arbitrary $n$ in string theory
due to the infinite number of particles in the string spectrum. This makes
the calculation of $n$-point string scattering amplitudes ($SSA$) with $%
n\geq5$ much more complicated. Indeed, as was shown by the present authors
recently that \cite{LLY2,Group,solve,LSSA} only $4$-point $SSA$ can be
expressed in terms of finite number of terms of Lauricella $SSA$ ($LSSA$).
Higher $n$-point $SSA$ with $n\geq5$ contain infinite number of terms of $%
LSSA$ \cite{long}.

On the other hand, for the case of $4$-point $SSA$, it has long been known
that all $4$-point $\left\langle V_{1}V_{2}V_{3}V_{4}\right\rangle $ hard $%
SSA$ ($E\rightarrow\infty$, fixed $\phi$) of different string states at each
fixed mass level of $V_{j}$ ($j=1,2,3,4$) vertex share the same functional
form \cite{ChanLee,ChanLee2, CHLTY2,CHLTY1}. See the reviews \cite{review,
over}. That is, all $4$-point hard $SSA$ ($HSSA$) at each fixed mass level
are proportional to each other with \textit{constant} ratios \cite%
{GM,GM1,Gross,GrossManes} (independent of the scattering angle $\phi$, or
the deficit of the kinematics variable dim$\mathcal{M}=1-0=1$).

Indeed, the first stringy scaling gives ratios among $4$-point $HSSA$ at
each fixed mass level $M^{2}=2(N-1)$ (Conjectured by Gross \cite{Gross},
proved by \cite{ChanLee,ChanLee2,CHLTY2,CHLTY1})%
\begin{equation}
\frac{\mathcal{T}^{\left( N,2m,q\right) }}{\mathcal{T}^{\left( N,0,0\right) }%
}=\frac{\left( 2m\right) !}{m!}\left( \frac{-1}{2M}\right) ^{2m+q}.\text{%
(independent of\textbf{\ }}\phi\text{ !!).}  \label{222}
\end{equation}
In Eq.(\ref{222}) $\mathcal{T}^{\left( N,m,q\right) }$ is the $4$-point $%
HSSA $ of any string vertex $V_{j}$ with $j=1,3,4$ and $V_{2}$ is the high
energy state in Eq.(\ref{111}); while $\mathcal{T}^{\left( N,0,0\right) }$
is the $4 $-point $HSSA$ of any string vertex $V_{j}$ with $j=1,3,4$, and $%
V_{2}$ is the leading Regge trajectory string state at mass level $N$. Note
that we have omitted the tensor indice of $V_{j}$ with $j=1,3,4$ and keep
only those of $V_{2}$ in $\mathcal{T}^{\left( N,2m,q\right) }$.

Moreover, the present authors discovered recently that the reduction of both
the number of kinematics variable dependence on the ratios and the number of
independent $HSSA$ for the $4$-point $HSSA$ can be generalized to arbitrary $%
n$-point $HSSA$ with $n\geq5$ \cite{hard,Regge}. As an example, all $6$%
-point $HSSA$ with $5$ tachyons and $1$ high energy state at mass level $%
M^{2}=2(N-1)$%
\begin{equation}
\left\vert p_{1},p_{2},p_{3};2m,2q\right\rangle =\left(
\alpha_{-1}^{T_{1}}\right) ^{N+p_{1}}\left( \alpha_{-1}^{T_{2}}\right)
^{p_{2}}\left( \alpha_{-1}^{T_{3}}\right) ^{p_{3}}\left(
\alpha_{-1}^{L}\right) ^{2m}\left( \alpha_{-2}^{L}\right) ^{q}\left\vert
0;k\right\rangle
\end{equation}
where $p_{1}+p_{2}=-2(m+q)$ with three transverse directions $T_{1}$, $T_{2}$
and $T_{3}$ are related to each other and the ratios are \cite{hard}
\begin{equation}
\frac{\mathcal{T}^{\left( \left\{ p_{1},p_{2},p_{3}\right\} ,m,q\right) }}{%
\mathcal{T}^{\left( \left\{ 0,0,0\right\} ,0,0\right) }}=\frac{\left(
2m\right) !}{m!}\left( \frac{-1}{2M_{2}}\right) ^{2m+q}\left( \cos
\theta_{1}\right) ^{p_{1}}\left( \sin\theta_{1}\cos\theta_{2}\right)
^{p_{2}}\left( \sin\theta_{1}\sin\theta_{2}\right) ^{p_{3}}  \label{16}
\end{equation}
where the number of kinematics variables reduced from $8$ to $2$, $%
\theta_{1} $ and $\theta_{2}$, and dim$\mathcal{M}=8-2=6$, a generalization
of dim$\mathcal{M}=1-0=1$ in Eq.(\ref{222}).

These \textit{stringy scaling} behaviors are reminiscent of Bjorken scaling
\cite{bs} and the Callan-Gross relation \cite{cg} in deep inelastic
scattering of electron and proton in the quark-parton model of $QCD$ where,
to the leading order in energy, the two structure functions $%
W_{1}(Q^{2},\nu) $ and $W_{2}(Q^{2},\nu)$ scale, and become not functions of
$2$ kinematics variables $Q^{2}$ and $\nu$ independently but only of their
ratio $Q^{2}/\nu$. The number of independent kinematics variables thus
reduces from $2$ to $1$, or the deficit dim$\mathcal{M}=2-1=1$. That is, the
structure functions scale as \cite{bs}%
\begin{equation}
MW_{1}(Q^{2},\nu)\rightarrow F_{1}(x),\text{ \ \ }\nu
W_{2}(Q^{2},\nu)\rightarrow F_{2}(x)\text{ (dim}\mathcal{M}=1\text{)}
\label{ss}
\end{equation}
where $x$ is the Bjorken variable and $M$ is the proton mass. Moreover, due
to the spin-$\frac{1}{2}$ assumption of quark, Callan and Gross derived the
relation \cite{cg}%
\begin{equation}
2xF_{1}(x)=F_{2}(x).  \label{ll}
\end{equation}
One easily sees that Eq.(\ref{16}) is the stringy generalization of $QCD$
scaling in Eq.(\ref{ss}) and Eq.(\ref{ll}). The next interesting issue is
then to understand the possible next to leading order stringy scaling
violation, similar to the $QCD$ corrections of Bjorken scaling or Bjorken
scaling violation through GLAP equation \cite{GL,AP} or current algebra.

To compare and make an anology between the stringy scaling and the Bjorken
scaling, we give a table for the two behaviors:

\bigskip%
\begin{tabular}{|l|l|}
\hline
Bosonic string & QCD \\ \hline
$SL(K+3,C)$ & $SU(3)$ \\ \hline
UV soft (exponential fall-off) & Asymptotic freedom \\ \hline
Nambu-Goto string model & Quark-parton model \\ \hline
Stringy scaling & Bjorken scaling \\ \hline
\textbf{Stringy scaling loop expansion} (stringy scaling violation) & GLAP
Eq. (Bjorken scaling violation). \\ \hline
\end{tabular}
\bigskip

Note that it was shown recently that all $n$-point SSA ($n\geq4$) of the
open bosonic string theory can be expressed in terms of the Lauricella
functions and form representation of the exact $SL(K+3,C)$ symmetry group
\cite{LSSA}. To define the integer $K$, a subset of exact $4$-point $SSA$
with three tachyons and one arbitrary string states (Note that $SSA$ of
three tachyons and one arbitrary string states with polarizations orthogonal
to the scattering plane \textit{vanish}.)%
\begin{equation}
\left\vert r_{n}^{T},r_{m}^{P},r_{l}^{L}\right\rangle =\prod_{n>0}\left(
\alpha_{-n}^{T}\right) ^{r_{n}^{T}}\prod_{m>0}\left( \alpha_{-m}^{P}\right)
^{r_{m}^{P}}\prod_{l>0}\left( \alpha_{-l}^{L}\right) ^{r_{l}^{L}}|0,k\rangle
\label{state}
\end{equation}
where $e^{P}=\frac{1}{M_{2}}(E_{2},\mathrm{k}_{2},0)=\frac{k_{2}}{M_{2}}$ is
the momentum polarization, $e^{L}=\frac{1}{M_{2}}(\mathrm{k}_{2},E_{2},0)$
is the longitudinal polarization and $e^{T}=(0,0,1)$ is the transverse
polarization on the $\left( 2+1\right) $-dimensional scattering plane, can
be expressed in terms of the $D$-type Lauricella functions \cite{LLY2}. In
addition to the mass level $M^{2}=2(N-1)$ with%
\begin{equation}
N=\sum_{\substack{ n,m,l>0  \\ \{\text{ }r_{j}^{X}\neq0\}}}\left(
nr_{n}^{T}+mr_{m}^{P}+lr_{l}^{L}\right) ,  \label{NN}
\end{equation}
we define another important index $K$ for the state in Eq.(\ref{state})%
\begin{equation}
K=\sum_{\substack{ n,m,l>0  \\ \{\text{ }r_{j}^{X}\neq0\}}}\left(
n+m+l\right)  \label{KKK}
\end{equation}
where $X=\left( T,P,L\right) $ and we have put $%
r_{n}^{T}=r_{m}^{P}=r_{l}^{L}=1$ in Eq.(\ref{NN}) in the definition of $K$.
Intuitively, $K$ counts the number of variaty of the $\alpha_{-j}^{X}$
oscillators in Eq.(\ref{state}).

On the other hand, it is interesting to see that while the stringy scaling
behavior was recognized only very recently, historically, the Bjorken
scaling was proposed before the invention of the idea of parton model, and
the discovery of asymptotic freedom was also motivated by the proposal of
Bjorken scaling.

To uncover the issue once and for all, in this paper we propose a systematic
approximation scheme to calculate general string-tree level $n$-point $HSSA$
order by order. We will show that the \textit{stringy scaling loop expansion
(}$SSLE$) scheme we proposed corresponds to finite number of vacuum diagram
terms (even for $n\geq5$) at each order of scattering energy due to a vacuum
diagram contraint and a topological graph constraint. Comparing to the
traditional effective action calculation for each loop diagram with infinite
number of external legs in field theory, finite number of vacuum diagrams
without external legs are much more easier to deal with.

In addition, we give the vacuum diagram representation and its Feynman rules
for each term in the expansion of the $HSSA$. In general, there can be many
vacuum diagrams, connected and disconnected, corresponds to one term in the
expansion. In particular, we match coefficient of each term with sum of the
inverse symmetry factors \cite{Peskin} corresponding to all diagrams of the
term. As an application to extending our previous calculation of $n$-point
leading order stringy scaling behavior, we explicitly calculate some
examples of $4$-point next to leading order stringy scaling violation terms.

This paper is organized as following. In the next section, we begin with the
stringy scaling loop expansion of the $4$-point $HSSA$. We will calculate in
details the functional form and coefficient of each term in the expansion.
Moreover, we give Feynman rules of vacuum diagram representation for each
term in the expansion. In section III and IV, we generalize the calculation
to the $5$-point and general $n$-point $HSSA$ respectively. In section V, we
demonstrate explicitly how to draw all the vacuum diagram representation,
connected and disconnected, for each term of the expansion. In particular,
we will sum over the inverse \textit{symmetry factors} of all diagrams of
the term to consistently match with the coefficient of the term. In section
VI, we use the results of section II to calculate some examples of $4$-point
next to leading order stringy scaling violation terms.. A brief conclusion
is given in section VII.

\setcounter{equation}{0} \renewcommand{\theequation}{\arabic{section}.%
\arabic{equation}}

\section{Stringy scaling loop expansion of $4$-point Amplitudes}

It can be exprecitly demonstrated that \cite{long} the $t-u$ channel of all $%
4$-point $SSA$ with four arbitrary tensor states can be written as the
following integral form (after $SL(2,R)$ fixing) \cite{long}%
\begin{equation}
\mathcal{T}(\Lambda)=\int_{1}^{\infty}dx\mbox{ }u(x)e^{-\Lambda f(x)},
\label{11}
\end{equation}
where $\Lambda\equiv-(1,2)=-k_{1}\cdot k_{2}$ . Indeed for the 26D open
bosonic string, a general state at mass level $N$%
\begin{equation}
M^{2}=2\left( N-1\right) ,\text{ and }N=\sum_{r>0}rn_{r},
\end{equation}
is of the form%
\begin{equation}
\left\vert P\right\rangle =\prod_{r>0}\prod_{\sigma=1}^{n_{r}}\frac {%
\varepsilon_{r}^{\left( \sigma\right) }\cdot\alpha_{-r}}{\sqrt {%
n_{r}!r^{n_{r}}}}|0,k\rangle
\end{equation}
where $\varepsilon_{r}^{\left( \sigma\right) }$ are polarizations with $%
\sigma=1,\cdots n_{r}$ for each operator $\alpha_{-r}$. The corresponding
string vertex is%
\begin{equation}
V\left( k,z\right) =\prod_{r>0}\prod_{\sigma=1}^{n_{r}}\varepsilon
_{r}^{\left( \sigma\right) }\cdot\partial_{z}^{r}X\left( z\right) e^{ik\cdot
X\left( z\right) }.  \label{2.4}
\end{equation}

For 4-point amplitude $i=1,2,3,4$, let%
\begin{equation}
X_{i}=X\left( z_{i}\right) \text{ and }k_{i}=\left( E_{i},\vec{p}_{i}\right)
\text{ with }\sum k_{i}=\sum\vec{p}_{i}=0,
\end{equation}
and we define the Mandelstam variables as $s=-\left( k_{1}+k_{2}\right) ^{2}$%
, $t=-\left( k_{2}+k_{3}\right) ^{2}$. The $4$-point SSA with four general
string states can be calculated as%
\begin{align}
A & =\int dz_{2}\left\vert z_{13}z_{14}z_{34}\right\vert \left\langle
\prod_{i=1}^{4}V_{i}\left( k_{i},z_{i}\right) \right\rangle  \notag \\
& =\int dz_{2}\left\vert z_{13}z_{14}z_{34}\right\vert \left\langle
\prod_{i=1}^{4}\exp\left( ik_{i}\cdot X_{i}+i\sum_{r_{i}>0}\sum_{\sigma
_{i}=1}^{n_{r_{i}}}\varepsilon_{r_{i}}^{\left( \sigma_{i}\right)
}\cdot\partial_{i}^{r_{i}}X_{i}\right) \right\rangle _{m.l.}  \notag \\
& =\int dz_{2}\left\vert z_{13}z_{14}z_{34}\right\vert \left\vert
z_{12}\right\vert ^{k_{1}\cdot k_{2}}\left\vert z_{13}\right\vert
^{k_{1}\cdot k_{3}}\left\vert z_{14}\right\vert ^{k_{1}\cdot
k_{4}}\left\vert z_{23}\right\vert ^{k_{2}\cdot k_{3}}\left\vert
z_{24}\right\vert ^{k_{2}\cdot k_{4}}\left\vert z_{34}\right\vert
^{k_{3}\cdot k_{4}}  \notag \\
& \cdot\exp\left(
\sum_{r_{i}>0}\sum_{\sigma_{i}=1}^{n_{r_{i}}}\sum_{i}^{4}\sum_{j\neq i}\frac{%
-\varepsilon_{r_{i}}^{\left( \sigma_{i}\right) }\cdot k_{j}}{z_{ji}^{r_{i}}}%
+\sum_{r_{i},r_{j}>0}\sum_{\sigma_{i}=1}^{n_{r_{i}}}\sum_{%
\sigma_{j}=1}^{n_{r_{j}}}\sum_{i<j=2}^{4}\frac {-\varepsilon_{r_{i}}^{\left(
\sigma_{i}\right) }\varepsilon_{r_{j}}^{\left( \sigma_{j}\right) }}{%
z_{ji}^{r_{i}}z_{ij}^{r_{j}}}\right) _{m.l.}
\end{align}
where the lower label $m.l.$ means that we only keep multi-linear terms with
each polarization $\varepsilon_{r_{i}}^{\left( \sigma_{i}\right) }$. The
amplitude can be expressed as%
\begin{align}
A & =\int_{0}^{1}dz_{2}z_{2}^{k_{1}\cdot k_{2}}\left( 1-z_{2}\right)
^{k_{2}\cdot k_{3}}  \notag \\
& \lim_{z_{4}\rightarrow\infty}\cdot\sum_{\left\{
\varepsilon_{r_{i}}^{\left( \sigma_{i}\right) }\right\} }\left[
\prod\limits_{i=1}^{4}\prod\limits_{\left\{ r_{i},\sigma_{i}\right\} }\left(
\sum_{j\neq i}\frac{\varepsilon_{r_{i}}^{\left( \sigma_{i}\right) }\cdot
k_{j}}{z_{ji}^{r_{i}}}\right)
\cdot\prod\limits_{i<j=2}^{4}\prod\limits_{\left\{
r_{i},\sigma_{i};r_{j},\sigma_{j}\right\} }\frac{\varepsilon_{r_{i}}^{\left(
\sigma_{i}\right) }\varepsilon_{r_{j}}^{\left( \sigma_{j}\right) }}{%
z_{ji}^{r_{i}}z_{ij}^{r_{j}}}\right] _{z_{1}=0,z_{3}=1}  \label{line}
\end{align}
where\ the configurations $\left\{ \varepsilon_{r_{i}}^{\left( \sigma
_{i}\right) }\right\} $\ satisfy%
\begin{equation}
\prod\limits_{i=1}^{4}\prod\limits_{\left\{ r_{i},\sigma_{i}\right\}
}\varepsilon_{r_{i}}^{\left( \sigma_{i}\right)
}\cdot\prod\limits_{i<j=2}^{4}\prod\limits_{\left\{
r_{i},\sigma_{i};r_{j},\sigma_{j}\right\} }\left(
\varepsilon_{r_{i}}^{\left( \sigma_{i}\right) }\varepsilon_{r_{j}}^{\left(
\sigma_{j}\right) }\right) =\prod_{i=1}^{4}\prod_{r_{i}>0}\prod_{\sigma
_{i}=1}^{n_{r_{i}}}\varepsilon_{r_{i}}^{\left( \sigma_{i}\right) },
\end{equation}
which ensures the multi-linear condition. For each configuration $\left\{
\varepsilon_{r_{i}}^{\left( \sigma_{i}\right) }\right\} $, it is
straightforward to transform Eq.(\ref{line}) to the standard integral form
in Eq.(\ref{11}).

\bigskip As a simple example, the $HSSA$ of three tachyons and one high
energy state $v_{2}$ \cite{CHLTY2,CHLTY1}%
\begin{equation}
\left\vert N,2m,q\right\rangle =\left( \alpha_{-1}^{T}\right)
^{N-2m-2q}\left( \alpha_{-1}^{L}\right) ^{2m}\left( \alpha_{-2}^{L}\right)
^{q}\left\vert 0;k\right\rangle  \label{111}
\end{equation}
with $e^{P}=\frac{1}{M_{2}}(E_{2},\mathrm{k}_{2},\vec{0})=\frac{k_{2}}{M_{2}}
$ the momentum polarization, $e^{L}=\frac{1}{M_{2}}(\mathrm{k}_{2},E_{2},%
\vec {0})$ the longitudinal polarization and the transverse polarization $%
e^{T}=(0,0,1)$ can be written as \cite{CHLTY2,CHLTY1}
\begin{align}
\mathcal{T}^{(N,2m,q)} & =\int_{1}^{\infty}dxx^{(1,2)}(1-x)^{(2,3)}\left[
\frac{e^{T}\cdot k_{1}}{x}-\frac{e^{T}\cdot k_{3}}{1-x}\right] ^{N-2m-2q}
\notag \\
& \cdot\left[ \frac{e^{P}\cdot k_{1}}{x}-\frac{e^{P}\cdot k_{3}}{1-x}\right]
^{2m}\left[ -\frac{e^{P}\cdot k_{1}}{x^{2}}-\frac{e^{P}\cdot k_{3}}{(1-x)^{2}%
}\right] ^{q},  \label{333}
\end{align}
which can then be put into the form in Eq.(\ref{11}) with
\begin{align}
\Lambda & \equiv-(1,2)\rightarrow\frac{s}{2}\rightarrow2E^{2}, \\
\tau & \equiv-\frac{(2,3)}{(1,2)}\rightarrow-\frac{t}{s}\rightarrow\sin ^{2}%
\frac{\phi}{2}, \\
f(x) & \equiv\ln x-\tau\ln(1-x), \\
u(x) & \equiv\left[ \frac{(1,2)}{M_{2}}\right] ^{2m+q}(1-x)^{-N+2m+2q}(f^{%
\prime})^{2m}(f^{\prime\prime})^{q}(-e^{T}\cdot k_{3})^{N-2m-2q}.
\end{align}

Note that to achieve BRST invariance or physical state conditions in the old
covariant quantization scheme for the state in the \textit{operator state
basis} ($OSB$) in Eq.(\ref{111}), one needs to add polarizations and put on
the Virasoro constraints. As an example, let's calculate the case of
symmetric spin $3$ state of mass level $M_{2}^{2}=4$. We first note that the
three momentum polarizations defined on the scattering plane above satisfy
the completeness relation

\begin{equation}
\eta^{\mu\nu}=\underset{\alpha,\beta}{\sum }e_{\alpha}^{\mu}e_{\beta}^{\nu}%
\eta^{\alpha\beta}  \label{2.11}
\end{equation}
where $\mu,\nu=0,1,2$ and $\alpha,\beta=P,L,T.$ $Diag$ $\eta^{\mu\nu
}=(-1,1,1)$. We can use Eq.(\ref{2.11}) to transform all $\mu,\nu$
coordinates to coordinates $\alpha,\beta$ on the scattering plane. One gauge
choice of the symmetric spin $3$ state in the\textit{\ physical state basis}
($PSB$) with Virasoro constraints can be calculated to be

\begin{equation}
\epsilon_{\mu\nu\lambda}\alpha_{-1}^{\mu\nu\lambda}\left\vert
0,k\right\rangle
;k^{\mu}\epsilon_{\mu\nu\lambda}=0,\eta^{\mu\nu}\epsilon_{\mu\nu\lambda}=0.
\label{2.22..}
\end{equation}
We can then use the helicity decomposition and writing $\epsilon_{\mu
\nu\lambda}=\Sigma_{\mu,\nu,\lambda}e_{\mu}^{\alpha}e_{\nu}^{\beta}e_{%
\lambda }^{\delta}u_{\alpha\beta\delta};\alpha,\beta,\delta=P,L,T$ to get

\begin{equation}
\epsilon_{\mu\nu\lambda}\alpha_{-1}^{\mu\nu\lambda}\left\vert
0,k\right\rangle
=[u_{TTL}(3\alpha_{-1}^{TTL}-\alpha_{-1}^{LLL})+u_{TTT}(\alpha_{-1}^{TTT}-3%
\alpha_{-1}^{LLT})]\left\vert 0,k\right\rangle .  \label{2.23}
\end{equation}
It is now easy to see from Eq.(\ref{2.23}) that to achieve BRST invariance,
the spin $3$ state in the\textit{\ }$PSB$\textit{\ }can be written as a
linear combination of states in the in the $OSB$\textit{\ }in Eq.(\ref{111})
with coefficients $u_{TTL}$ and $u_{TTT}$. Similar procedure can be
performed to achieve BRST invariance of states in Eq.(\ref{state}) and Eq.(%
\ref{2.4}).

In general for four arbitrary string states, we can expand the amplitude\ in
Eq.(\ref{11}) around the saddle point for large $\Lambda$ to obtain

\begin{align}
\mathcal{T}(\Lambda) & =\int_{1}^{\infty}dx\text{ }u\left( x\right)
e^{-\Lambda f\left( x\right) }  \notag \\
& =\int_{1}^{\infty}dx\left( \sum_{p\geq0}\frac{u_{0}^{\left( p\right) }}{p!}%
\left( x-x_{0}\right) ^{p}\right) e^{-\Lambda f_{0}-\frac{1}{2}\Lambda
f_{0}^{\prime\prime}\left( x-x_{0}\right) ^{2}-\Lambda\sum_{j=3}\frac {%
f_{0}^{\left( j\right) }}{j!}\left( x-x_{0}\right) ^{j}}  \notag \\
& =e^{-\Lambda f_{0}}\int_{1}^{\infty}dx\left( \sum_{p\geq0}\sum_{q\geq 0}%
\frac{\left( -\Lambda\right) ^{q}u_{0}^{\left( p\right) }}{i!p!}\left(
x-x_{0}\right) ^{p}\left[ \sum_{j\geq3}\frac{f_{0}^{\left( j\right) }}{j!}%
\left( x-x_{0}\right) ^{j}\right] ^{q}\right) e^{-\frac{1}{2}\Lambda
f_{0}^{\prime\prime}\left( x-x_{0}\right) ^{2}}.  \label{tt}
\end{align}
Let's rewrite the bracket term in the last line of the above equation as%
\begin{align}
\left[ \sum_{j\geq3}\frac{f_{0}^{\left( j\right) }}{j!}\left( x-x_{0}\right)
^{j}\right] ^{q} & =\left[ \sum_{n_{1}\geq3}a_{n_{1}}z^{n_{1}}\right] \left[
\sum_{n_{2}\geq3}a_{n_{2}}z^{n_{2}}\right] \cdots\left[ \sum_{n_{q}%
\geq3}a_{n_{q}}z^{n_{q}}\right]  \notag \\
& =\sum_{n_{1},\cdots n_{q}\geq3}a_{n_{1}}\cdots
a_{n_{q}}z^{\sum_{r=1}^{q}n_{r}}=\sum_{n_{1},\cdots
n_{q}\geq0}a_{n_{1}+3}\cdots a_{n_{q}+3}z^{\sum_{r=1}^{q}(n_{r}+3)}.
\label{ii}
\end{align}
Inserting Eq.(\ref{ii}) into Eq.(\ref{tt}), and using the Gaussian integral%
\begin{equation}
\int_{-\infty}^{\infty}dzz^{2n}e^{-\frac{1}{2}z^{2}}=\sqrt{2\pi}\frac {(2n)!%
}{2^{n}n!}  \label{g}
\end{equation}
to perform the integration, we obtain%
\begin{align}
\mathcal{T}(\Lambda) & =\sqrt{\frac{2\pi}{\Lambda f_{0}^{\prime\prime}}}%
e^{-\Lambda f_{0}}\sum_{q\geq0}\sum_{n_{1},\cdots n_{q}\geq0}\sum_{p\geq 0}%
\frac{(-)^{q}(2M+2q)!}{q!2^{M+q}(M+q)!}  \notag \\
& \cdot\frac{u_{0}^{(p)}f_{0}^{(n_{1}+3)}\cdots f_{0}^{(n_{q}+3)}}{%
p!(n_{1}+3)!\cdots(n_{q}+3)!(f_{0}^{\prime\prime})^{M+q}}\frac{1}{\Lambda
^{M}}  \notag \\
& \equiv\sqrt{\frac{2\pi}{\Lambda f_{0}^{\prime\prime}}}e^{-\Lambda f_{0}}%
\left[ \mathcal{A}(\Lambda^{0})+\frac{1}{\Lambda}\mathcal{A}(\Lambda ^{-1})+%
\frac{1}{\Lambda^{2}}\mathcal{A}(\Lambda^{-2})+O\left( \frac {1}{\Lambda^{M}}%
\right) \right]  \label{main}
\end{align}
where%
\begin{equation}
2M=p+\sum_{r=1}^{q}(n_{r}+1)\geq0.  \label{con1}
\end{equation}
In Eq.(\ref{con1}), $M$, $p$, $q$ and $n_{r}$ are nonnegative integers. It
is important to note that for a given inverse energy order $\frac{1}{%
\Lambda^{M}}$, there are only \textit{finite} number of terms in Eq.(\ref%
{main}) due to the condition in Eq.(\ref{con1}). We can now explicitly
calculate $\mathcal{A}(\Lambda)$ in Eq.(\ref{main}) order by order.

For the leading order $M=0$, we have $p=0$, $q=0$ and there is no $n_{r}$.
The amplitude is%
\begin{equation}
\mathcal{A}(\Lambda^{0})=u_{0}.  \label{00}
\end{equation}
\bigskip For the next to leading order $M=1$, there are $4$ terms:%
\begin{align}
\mathcal{A}_{1}(\Lambda^{-1}) & =-\frac{u_{0}f_{0}^{(4)}}{%
8(f_{0}^{\prime\prime})^{2}},\text{\ }(p=0,q=1,n_{1}=1)  \label{1111} \\
\mathcal{A}_{2}(\Lambda^{-1}) & =\frac{5u_{0}(f_{0}^{(3)})^{2}}{%
24(f_{0}^{\prime\prime})^{3}},\text{ \ }(p=0,q=2,n_{1}=n_{2}=0)  \label{5555}
\\
\mathcal{A}_{3}(\Lambda^{-1}) & =-\frac{u_{0}^{\prime}f_{0}^{(3)}}{%
2(f_{0}^{\prime\prime})^{2}},\text{\ }(p=1,q=1,n_{1}=0)  \label{6666} \\
\mathcal{A}_{4}(\Lambda^{-1}) & =\frac{u_{0}^{\prime\prime}}{%
2f_{0}^{\prime\prime}}.\text{\ }(p=2,q=0)  \label{4444}
\end{align}
For the next next to leading order $M=2$, there are $12$ terms:%
\begin{align}
\mathcal{A}_{1}(\Lambda^{-2}) & =-\frac{u_{0}f_{0}^{(6)}}{%
48(f_{0}^{\prime\prime})^{3}},\text{\ }(p=0,q=1,n_{1}=3)  \label{2222} \\
\mathcal{A}_{2}(\Lambda^{-2}) & =\frac{7u_{0}f_{0}^{(3)}f_{0}^{(5)}}{%
48(f_{0}^{\prime\prime})^{4}},\text{ \ }(p=0,q=2,(n_{1},n_{2})=(2,0)\text{
or }(0,2)) \\
\mathcal{A}_{3}(\Lambda^{-2}) & =\frac{35u_{0}(f_{0}^{(4)})^{2}}{%
384(f_{0}^{\prime\prime})^{4}},\text{\ }(p=0,q=2,n_{1}=n_{2}=1) \\
\mathcal{A}_{4}(\Lambda^{-2}) & =-\frac{35u_{0}f_{0}^{(4)}(f_{0}^{(3)})^{2}}{%
64(f_{0}^{\prime\prime})^{5}},\text{\ }(p=0,q=3,(n_{1},n_{2},n_{3})=(1,0,0)%
\text{ or permutation}) \\
\mathcal{A}_{5}(\Lambda^{-2}) & =\frac{385u_{0}(f_{0}^{(3)})^{4}}{%
1152(f_{0}^{\prime\prime})^{6}},\text{\ }(p=0,q=4,n_{1}=n_{2}=n_{3}=n_{4}=0)
\\
\mathcal{A}_{6}(\Lambda^{-2}) & =-\frac{u_{0}^{^{\prime}}f_{0}^{(5)}}{%
8(f_{0}^{\prime\prime})^{3}},\text{\ }(p=1,q=1,n_{1}=2) \\
\mathcal{A}_{7}(\Lambda^{-2}) & =\frac{35u_{0}^{^{%
\prime}}f_{0}^{(3)}f_{0}^{(4)}}{48(f_{0}^{\prime\prime})^{4}},\text{\ }%
(p=1,q=2,(n_{1},n_{2})=(1,0)\text{ or }(0,1)) \\
\mathcal{A}_{8}(\Lambda^{-2}) & =-\frac{35u_{0}^{^{\prime}}(f_{0}^{(3)})^{2}%
}{48(f_{0}^{\prime\prime})^{5}},\text{\ }(p=1,q=3,n_{1}=n_{2}=n_{3}=0) \\
\mathcal{A}_{9}(\Lambda^{-2}) & =-\frac{5u_{0}^{^{\prime\prime}}f_{0}^{(4)}}{%
16(f_{0}^{\prime\prime})^{3}},\text{\ }(p=2,q=1,n_{1}=1) \\
\mathcal{A}_{10}(\Lambda^{-2}) & =\frac{35u_{0}^{\prime%
\prime}(f_{0}^{(3)})^{2}}{48(f_{0}^{\prime\prime})^{4}},\text{\ }%
(p=2,q=2,n_{1}=n_{2}=0) \\
\mathcal{A}_{11}(\Lambda^{-2}) & =-\frac{5u_{0}^{(3)}f_{0}^{(3)}}{%
12(f_{0}^{\prime\prime})^{3}},\text{\ }(p=3,q=1,n_{1}=0) \\
\mathcal{A}_{12}(\Lambda^{-2}) & =\frac{u_{0}^{(4)}}{8(f_{0}^{\prime\prime
})^{2}}\text{\ }.(p=4,q=0,\text{ No }n_{r})  \label{3333}
\end{align}

To study the general higher order amplitudes, in addition to Eq.(\ref{main}%
), we calculate an alternative expansion of the amplitudes which is suitable
to diagrammatic representations to be discussed later. First, we perform a
Taylor expansion of $u$ and $f$ at the saddle point where the first
derivative of $f$ , $f_{0}^{\left( 1\right) }$ is zero

\begin{align}
\mathcal{T}(\Lambda )& =\int dx\text{ }u\left( x\right) e^{-\Lambda f\left(
x\right) }  \notag \\
& =\int dx\left( \sum_{m\geq 0}\frac{u_{0}^{\left( m\right) }}{m!}\left(
x-x_{0}\right) ^{m}\right) e^{-\Lambda f_{0}-\frac{1}{2}\Lambda
f_{0}^{\prime \prime }\left( x-x_{0}\right) ^{2}-\Lambda \sum_{n=3}\frac{%
f_{0}^{\left( n\right) }}{n!}\left( x-x_{0}\right) ^{n}}  \notag \\
& =\int dx\left( \sum_{m\geq 0}\frac{u_{0}^{\left( m\right) }}{m!}\left(
x-x_{0}\right) ^{m}\right) e^{-\Lambda f_{0}-\frac{1}{2}\Lambda
f_{0}^{\prime \prime }\left( x-x_{0}\right) ^{2}}\exp \left( -\Lambda
\sum_{n=3}\frac{f_{0}^{\left( n\right) }}{n!}\left( x-x_{0}\right)
^{n}\right) .
\end{align}%
The next step is to expand the terms of the exponential function starting
from the third derivative.%
\begin{equation}
\mathcal{T}(\Lambda )=\int dx\left( \sum_{m\geq 0}\frac{u_{0}^{\left(
m\right) }}{m!}\left( x-x_{0}\right) ^{m}\right) e^{-\Lambda f_{0}-\frac{1}{2%
}\Lambda f_{0}^{\prime \prime }\left( x-x_{0}\right) ^{2}}\sum\limits_{L=0}%
\frac{1}{L!}\left( -\Lambda \sum_{n=3}\frac{f_{0}^{\left( n\right) }}{n!}%
\left( x-x_{0}\right) ^{n}\right) ^{L}
\end{equation}%
where we have used the multinomial theorem to expand the $L$-th power%
\begin{equation}
\left( -\Lambda \sum_{n=3}\frac{f_{0}^{\left( n\right) }}{n!}\left(
x-x_{0}\right) ^{n}\right) ^{L}=\frac{L!}{\prod\limits_{n\geq 3}V\left(
n\right) !}\prod\limits_{n\geq 3}\left( \frac{-\Lambda f_{0}^{\left(
n\right) }}{n!}\left( x-x_{0}\right) ^{n}\right) ^{V\left( n\right) }
\end{equation}%
to obtain%
\begin{equation}
\mathcal{T}(\Lambda )=\int dx\left( \sum_{m\geq 0}\frac{u_{0}^{\left(
m\right) }}{m!}\left( x-x_{0}\right) ^{m}\right) e^{-\Lambda f_{0}-\frac{1}{2%
}\Lambda f_{0}^{\prime \prime }\left( x-x_{0}\right) ^{2}}\sum_{\left\{
V\left( n\right) \right\} }\left( \prod\limits_{n\geq 3}\frac{1}{V\left(
n\right) !}\left( \frac{-\Lambda f_{0}^{\left( n\right) }}{n!}\right)
^{V\left( n\right) }\left( x-x_{0}\right) ^{nV\left( n\right) }\right) .
\end{equation}%
The $\left\{ V\left( n\right) \right\} $ symbol of the above equation
denotes the integer partitions of $L$ into positive integers and%
\begin{equation}
L=\sum\limits_{n\geq 3}V\left( n\right) .
\end{equation}%
\qquad

We can now use Eq.(\ref{g}) to perform the following integration%
\begin{equation}
\mathcal{T}(\Lambda )=\sum_{m\geq 0}\frac{u_{0}^{\left( m\right) }}{m!}%
\sum_{\left\{ V\left( n\right) \right\} }\left[ \left( \prod\limits_{n\geq 3}%
\frac{1}{V\left( n\right) !}\left( \frac{-\Lambda f_{0}^{\left( n\right) }}{%
n!}\right) ^{V\left( n\right) }\right) \int dxe^{-\Lambda f_{0}-\frac{1}{2}%
\Lambda f_{0}^{\prime \prime }\left( x-x_{0}\right) ^{2}}\left(
x-x_{0}\right) ^{m+\sum_{n=3}nV\left( n\right) }\right]
\end{equation}%
to get
\begin{equation}
\mathcal{T}(\Lambda )=\sum_{m\geq 0}\frac{u_{0}^{\left( m\right) }}{m!}%
\sum_{\left\{ V\left( n\right) \right\} }\left( \prod\limits_{n\geq 3}\frac{1%
}{V\left( n\right) !}\left( \frac{-\Lambda f_{0}^{\left( n\right) }}{n!}%
\right) ^{V\left( n\right) }\sqrt{\frac{2\pi }{\Lambda f_{0}^{\prime \prime }%
}}\frac{1}{\left( \Lambda f_{0}^{\prime \prime }\right) ^{P}}e^{-\Lambda
f_{0}}\frac{(2P)!}{2^{P}P!}\right)
\end{equation}%
where we have defined $m+\sum_{n=3}nV\left( n\right) =2P$ and note that the
integral is nonzero only when $m+\sum_{n=3}nV\left( n\right) $ is even.

Finally we define $P-\sum_{n=3}V\left( n\right) =M$ to count the order of $%
\Lambda $ and obtain
\begin{eqnarray}
\mathcal{T}(\Lambda ) &=&\sqrt{\frac{2\pi }{\Lambda f_{0}^{\prime \prime }}}%
e^{-\Lambda f_{0}}\sum_{m\geq 0}\sum_{\left\{ V\left( n\right) \right\} }%
\frac{1}{\Lambda ^{M}}\frac{(2P)!}{2^{P}P!\left( f_{0}^{\prime \prime
}\right) ^{P}}\frac{u_{0}^{\left( m\right) }}{m!}\prod\limits_{n\geq
3}\left( \frac{\left( f_{0}^{\left( n\right) }\right) ^{V\left( n\right) }}{%
\left( -n!\right) ^{V\left( n\right) }V\left( n\right) !}\right)
\label{main2} \\
&=&\sqrt{\frac{2\pi }{\Lambda f_{0}^{\prime \prime }}}e^{-\Lambda f_{0}}%
\left[ \mathcal{A}(\Lambda ^{0})+\frac{1}{\Lambda }\mathcal{A}(\Lambda
^{-1})+\frac{1}{\Lambda ^{2}}\mathcal{A}(\Lambda ^{-2})+O\left( \frac{1}{%
\Lambda ^{M}}\right) \right] .
\end{eqnarray}%
The above expansion is subject to the following conditions%
\begin{eqnarray}
m+\sum_{n=3}nV\left( n\right) &=&2P, \\
P-\sum_{n=3}V\left( n\right) &=&M
\end{eqnarray}%
where $P>0$ and $M\geq 0$.

We note that a typical term at each order $\Lambda ^{-M}$ in the expansion
of Eq.(\ref{main2}) can be written as%
\begin{equation}
\mathcal{A}(\Lambda ^{-M})\sim \frac{\left( 2P\right) !}{P!2^{P}}\frac{1}{m!}%
\left[ \underset{n\geq 3}{\prod }\frac{1}{\left( -n!\right) ^{V\left(
n\right) }V\left( n\right) !}\right] \cdot \frac{u_{0}^{\left( m\right) }%
\underset{n\geq 3}{\prod }\left( f_{0}^{\left( n\right) }\right) ^{V\left(
n\right) }}{\left( f_{0}^{\left( 2\right) }\right) ^{P}}.  \label{sum1}
\end{equation}%
The rules (corresponding to symmetry factors of Feynman rules in field
theory, see section V for more details) to assign constant factors in the
bracket of Eq.(\ref{sum1}) are

\begin{align}
u_{0}^{\left( m\right) } & \Rightarrow\frac{1}{m!}, \\
\underset{n\geq3}{\prod}\left( f_{0}^{\left( n\right) }\right) ^{V\left(
n\right) } & \Rightarrow\underset{n\geq3}{\prod}\frac{1}{\left( -n!\right)
^{V\left( n\right) }V\left( n\right) !}, \\
\left( f_{0}^{\left( 2\right) }\right) ^{-P} & \Rightarrow\frac{\left(
2P\right) !}{P!2^{P}}=\left( 2P-1\right) !!.  \label{3c}
\end{align}
Note that the factor in Eq.(\ref{3c}) can be interpreted as the coefficient
of $x_{2}^{P}$ term in the expansion of the incomplete Bell polynomials $%
B_{n,k}$ $(x_{1},x_{2},\cdots,x_{n-k+1})$ with $n=2P$ and $k=P$ since there
are $P$ propagators each with $2$ end points. We have verified coefficients
of all terms in Eq.(\ref{1111}) to Eq.(\ref{4444}) calculated previously in $%
\mathcal{A}_{j}(\Lambda^{-1})$ and all terms in Eq.(\ref{2222}) to Eq.(\ref%
{3333}) calculated in $\mathcal{A}_{j}(\Lambda^{-2})$ by using Eq.(\ref{sum1}%
).

It is remarkable that each typical term in Eq.(\ref{sum1}) corresponds to
(at least) one vacuum Feynman diagram (no external legs). Here we list the
rules regarding the expansion and the construction of a vacuum diagram
corresponds to the typical term in Eq.(\ref{sum1}):

\begin{equation}
\begin{tabular}{l}
$V\left( n\right) $ $n$-vertex $\sim\left( f_{0}^{\left( n\right) }\right)
^{V\left( n\right) }$ for $n\geq3$, \\
$P$ propagators $\sim\left( f_{0}^{\prime\prime}\right) ^{P}$, \\
a loop with $m$ legs $\sim$ $u_{0}^{\left( m\right) }$ ( if $m=0$, $u_{0}$
will be treated as a disconnected loop), \\
$M=$ \# of loops $-$ \# of the connected components $\geq1$.%
\end{tabular}
\   \label{FR1}
\end{equation}

Note that some terms in Eq.(\ref{sum1}) can correspond to more than one
diagram. However, for each order of $M$, there are only finite number of
terms (diagrams) in the stringy scaling loop expansion scheme.

The constraints for the parameters are%
\begin{align}
P-\sum_{n=3}V\left( n\right) & =M\text{,}  \label{55} \\
m+\sum_{n=3}nV\left( n\right) & =2P\Rightarrow\text{vacuum diagram.}
\label{66}
\end{align}
Note that Eq.(\ref{55}) can be read from Eq.(\ref{main}), and Eq.(\ref{66})
is equivalent to Eq.(\ref{con1}). On the other hand, Eq.(\ref{55}) means
that $M$ is the difference between the number of $f_{0}^{\left( n\right) }$
in the numerator and the number of $f_{0}^{(2)}$ in the denominator, and Eq.(%
\ref{66}) means that the number of differentiations of $f$ in the numerator
equals to the number of differentiations in the denominator. We will see
that Eq.(\ref{55}) and Eq.(\ref{66}) give a vacuum diagram representation
for each term in Eq.(\ref{sum1}). While Eq.(\ref{66}) gives the vacuum
diagram condition, topologically, Eq.(\ref{55}) follows from the Euler
characteristics $\chi(\mathbb{M})$ with dim$\mathbb{M}=1$%
\begin{equation}
\chi(\mathbb{M})=\sum_{n=3}V\left( n\right) -P=-M
\end{equation}
where the number of the $n$-vertex is $V\left( n\right) $, the number of
edges is $P$ and the number of faces of the $1D$ graph manifold $\mathbb{M}$
is zero. Indeed, for this case, the Euler characteristics can also be
written as%
\begin{equation}
\chi(\mathbb{M})=b_{0}-b_{1}=-M
\end{equation}
where $b_{j}$ is the $j$th Betti number of $\mathbb{M}$. Here $b_{0}$ counts
the number of the connected components of the diagram and $b_{1}$ counts the
total number of loops of the diagram.

Eliminating $P$ from the above constraints Eq.(\ref{55}) and Eq.(\ref{66}),
we obtain the following equation%
\begin{equation}
\sum_{n=3}^{2l}\left( n-2\right) V\left( n\right) =2M-m\geq0.  \label{sol}
\end{equation}
For a given integer $M\geq1$,%
\begin{equation}
m=0,1,\cdots,2M.
\end{equation}

One can solve all non-negative integer solutions for $V\left( n\right) $
with $n\geq3$ in Eq.(\ref{sol}).

For the $\Lambda^{-1}$ order, i.e. $M=1$, we get%
\begin{equation}
\begin{tabular}{|c|c|c|c|c|}
\hline
$m$ & $0$ & $0$ & $1$ & $2$ \\ \hline
$V\left( 3\right) $ & $0$ & $2$ & $1$ & $0$ \\ \hline
$V\left( 4\right) $ & $1$ & $0$ & $0$ & $0$ \\ \hline
\end{tabular}
\Rightarrow4\text{ terms}
\end{equation}
as expected from the previous calculation.

For the $\Lambda^{-2}$ order, i.e.$M=2$, we get%
\begin{equation}
\begin{tabular}{|c|c|c|c|c|c|c|c|c|c|c|c|c|}
\hline
$m$ & $0$ & $0$ & $0$ & $0$ & $0$ & $1$ & $1$ & $1$ & $2$ & $2$ & $3$ & $4$
\\ \hline
$V\left( 3\right) $ & $0$ & $4$ & $1$ & $2$ & $0$ & $3$ & $0$ & $1$ & $2$ & $%
0$ & $1$ & $0$ \\ \hline
$V\left( 4\right) $ & $0$ & $0$ & $0$ & $1$ & $2$ & $0$ & $0$ & $1$ & $0$ & $%
1$ & $0$ & $0$ \\ \hline
$V\left( 5\right) $ & $0$ & $0$ & $1$ & $0$ & $0$ & $0$ & $1$ & $0$ & $0$ & $%
0$ & $0$ & $0$ \\ \hline
$V\left( 6\right) $ & $1$ & $0$ & $0$ & $0$ & $0$ & $0$ & $0$ & $0$ & $0$ & $%
0$ & $0$ & $0$ \\ \hline
\end{tabular}
\Rightarrow12\text{ terms}
\end{equation}
as expected from the previous calculation.

For the higher order amplitudes, the total number of terms are
\begin{equation}
\begin{tabular}{|c|c|c|c|c|c|c|c|c|c|c|}
\hline
$M$ & $1$ & $2$ & $3$ & $4$ & $5$ & $6$ & $7$ & $8$ & $9$ & $\cdots$ \\
\hline
$\#$ of terms & $4$ & $12$ & $30$ & $67$ & $139$ & $272$ & $508$ & $915$ & $%
1597$ & $\cdots$ \\ \hline
\end{tabular}
.
\end{equation}
On the other hand, for a given $M$, we can count the number of terms for
each $m$%
\begin{equation}
\begin{tabular}{|c|c|c|c|c|c|c|c|c|c|c|c|c|}
\hline
$m$ & $0$ & $1$ & $2$ & $3$ & $4$ & $5$ & $6$ & $7$ & $8$ & $9$ & $10$ &
total \\ \hline
$M=1$ & $2$ & $1$ & $1$ &  &  &  &  &  &  &  &  & $4$ \\ \hline
$M=2$ & $5$ & $3$ & $2$ & $1$ & $1$ &  &  &  &  &  &  & $12$ \\ \hline
$M=3$ & $11$ & $7$ & $5$ & $3$ & $2$ & $1$ & $1$ &  &  &  &  & $30$ \\ \hline
$M=4$ & $22$ & $15$ & $11$ & $7$ & $5$ & $3$ & $2$ & $1$ & $1$ &  &  & $67$
\\ \hline
$M=5$ & $42$ & $30$ & $22$ & $15$ & $11$ & $7$ & $5$ & $3$ & $2$ & $1$ & $1$
& $139$ \\ \hline
\end{tabular}
\ \ \ .  \label{terms}
\end{equation}
We observe from the above table that the distribution on $m$ for a given $M$%
\begin{equation}
1,1,2,3,5,7,11,15,22,30,42,\cdots
\end{equation}
can be generated by the generating function%
\begin{align}
\prod\limits_{n=1}^{\infty}\left( 1-q^{n}\right) ^{-1} &
=1+q+2q^{2}+3q^{3}+5q^{4}+7q^{5}+11q^{6}+15q^{7}  \notag \\
& +22q^{8}+30q^{9}+42q^{10}+56q^{11}+77q^{12}+\cdots  \notag \\
& =\sum_{n=0}^{\infty}P(n)q^{n}  \label{p}
\end{align}
which is the inversed Dedekind eta function. It corresponds to the scalar
partition function on the torus containing the information of the number of
states at each energy level or character of a conformal family. $P(n)$ in
Eq.(\ref{p}) is the number of ways of writing $n$ as a sum of positive
integer. From Eq.(\ref{sol}), we easily see that the numer of terms $%
N_{(M,m)}$ for given $M$ and $m$ presented in Eq.(\ref{terms}) is%
\begin{equation}
N_{(M,m)}=P(2M-m).  \label{N}
\end{equation}

\setcounter{equation}{0} \renewcommand{\theequation}{\arabic{section}.%
\arabic{equation}}

\section{Stringy scaling loop expansion of $5$-point Amplitudes}

The $5$-point $SSA$ can be written in the following integral form (after $%
SL(2,R)$ fixing)%
\begin{equation}
\mathcal{T}(\Lambda)=\int dx_{2}dx_{3}\text{ }u\left( x_{2},x_{3}\right)
e^{-\Lambda f\left( x_{2},x_{3}\right) },\text{ }\Lambda=-k_{1}\cdot k_{2},
\label{T5}
\end{equation}
where%
\begin{equation}
f\left( x_{2},x_{3}\right) =-\frac{k_{1}\cdot k_{2}}{\Lambda}\ln x_{2}-\frac{%
k_{1}\cdot k_{3}}{\Lambda}\ln x_{3}-\frac{k_{2}\cdot k_{3}}{\Lambda}%
\ln\left( x_{3}-x_{2}\right) -\frac{k_{2}\cdot k_{4}}{\Lambda}\ln\left(
1-x_{2}\right) -\frac{k_{3}\cdot k_{4}}{\Lambda}\ln\left( 1-x_{3}\right) .
\end{equation}
Since we are going to use the Gaussian approximation and perform the
integration of Eq.(\ref{T5}) by Eq.(\ref{g}), for the time being, we will
ignore the range of integration in Eq.(\ref{T5}).

As a simple example, for the $5$-point $HSSA$ with $4$ tachyons and $1$ high
energy state at mass level $M^{2}=2(N-1)$%
\begin{equation}
\left\vert p_{1},p_{2};2m,2q\right\rangle =\left( \alpha_{-1}^{T_{1}}\right)
^{N+p_{1}}\left( \alpha_{-1}^{T_{2}}\right) ^{p_{2}}\left(
\alpha_{-1}^{L}\right) ^{2m}\left( \alpha_{-2}^{L}\right) ^{q}\left\vert
0;k\right\rangle  \label{pr}
\end{equation}
where $p_{1}+p_{2}=-2(m+q)$ with two transverse directions $T_{1}$ and $%
T_{2} $, $u\left( x_{2},x_{3}\right) $ can be calculated to be

\begin{equation}
u\left( x_{2},x_{3}\right) =\left( k^{T_{1}}\right) ^{N+p_{1}}\left(
k^{T_{2}}\right) ^{p_{2}}\cdots\left( k^{T_{r}}\right) ^{p_{r}}\left(
k^{L}\right) ^{2m}\left( k^{\prime L}\right) ^{q},  \label{u5}
\end{equation}
where we have defined%
\begin{equation}
k=-\frac{k_{1}}{x_{1}-x_{2}}-\frac{k_{3}}{x_{3}-x_{2}}-\frac{k_{4}}{%
x_{4}-x_{2}}.  \label{kp}
\end{equation}
We perform Taylor expansions on the saddle point $\left(
x_{20},x_{30}\right) $ of both the $u$ and $f$ \ functions to obtain%
\begin{align}
\mathcal{T}(\Lambda) & =\int dx_{2}dx_{3}\text{ }\left[ u\left(
x_{20},x_{30}\right) +\cdots\right]  \notag \\
& \cdot e^{-\Lambda\left[ f\left( x_{20},x_{30}\right) +\frac{1}{2}\frac{%
\partial^{2}f\left( x_{2},x_{3}\right) }{\partial x_{2}^{2}}\left(
x_{2}-x_{20}\right) ^{2}+\frac{\partial^{2}f\left( x_{2},x_{3}\right) }{%
\partial x_{2}\partial x_{3}}\left( x_{2}-x_{20}\right) \left(
x_{3}-x_{30}\right) +\frac{\partial^{2}f\left( x_{2},x_{3}\right) }{\partial
x_{3}^{2}}\left( x_{3}-x_{30}\right) ^{2}+\cdots\right] }
\end{align}
where $\left( x_{20},x_{30}\right) $ satisfies%
\begin{equation}
\frac{\partial f\left( x_{20},x_{30}\right) }{\partial x_{2}}=0,\text{ \ \ }%
\frac{\partial f\left( x_{20},x_{30}\right) }{\partial x_{3}}=0.
\end{equation}

We observe that in the Taylor expansion of the function $f$ , there are
crossing terms such as $\frac{\partial^{2}f\left( x_{2},x_{3}\right) }{%
\partial x_{2}\partial x_{3}}$ which involves $(x_{2}-x_{20})(x_{3}-x_{30})$%
. These crossing terms will result in an infinite number of terms at each
order $\Lambda$ of expansion of $\mathcal{T}(\Lambda)$ in the limit as $%
\Lambda\rightarrow\infty$ . Therefore, we need to do a change of variables
here to eliminate these crossing terms and obtain
\begin{equation}
\mathcal{T}(\Lambda)=\int dx_{2}^{\prime}dx_{3}^{\prime}\text{ }\left[
u\left( x_{20}^{\prime},x_{30}^{\prime}\right) +\cdots\right] e^{-\Lambda%
\left[ f\left( x_{20}^{\prime},x_{30}^{\prime}\right) +\frac {1}{2}\frac{%
\partial^{2}f\left( x_{2}^{\prime},x_{3}^{\prime}\right) }{\partial
x_{2}^{\prime2}}\left( x_{2}^{\prime}-x_{20}^{\prime}\right) ^{2}+\frac{1}{2}%
\frac{\partial^{2}f\left( x_{2}^{\prime},x_{3}^{\prime }\right) }{\partial
x_{3}^{\prime2}}\left( x_{3}^{\prime}-x_{30}^{\prime }\right) ^{2}+\cdots%
\right] }
\end{equation}
where $\left( x_{20}^{\prime},x_{30}^{\prime}\right) $ satisfies%
\begin{equation}
\frac{\partial f\left( x_{20}^{\prime},x_{30}^{\prime}\right) }{\partial
x_{2}^{\prime}}=0,\text{ \ \ }\frac{\partial f\left(
x_{20}^{\prime},x_{30}^{\prime}\right) }{\partial x_{3}^{\prime}}=0.
\end{equation}
Let's define the coefficients in the Taylor expansion of $f$ and $u$ at $%
\left( x_{20}^{\prime},x_{30}^{\prime}\right) $ as follows:%
\begin{align}
u\left( x_{20}^{\prime},x_{30}^{\prime}\right) & =u_{0}, \\
\frac{\partial^{m_{2}+m_{3}}u\left( x_{20}^{\prime},x_{30}^{\prime}\right) }{%
\partial\left( x_{2}^{\prime}\right) ^{m_{2}}\partial\left(
x_{3}^{\prime}\right) ^{m_{3}}} & =u_{0}^{\left( m_{2},m_{3}\right) }, \\
\frac{\partial^{n_{2}+n_{3}}f\left( x_{20}^{\prime},x_{30}^{\prime}\right) }{%
\partial\left( x_{2}^{\prime}\right) ^{n_{2}}\partial\left(
x_{3}^{\prime}\right) ^{n_{3}}} & =f_{0}^{\left( n_{2},n_{3}\right) }.
\end{align}
We can then simplify the integral into the following form
\begin{equation}
\mathcal{T}(\Lambda)=\int dx_{2}^{\prime}dx_{3}^{\prime}\text{ }\left[
u_{0}+\cdots\right] e^{-\Lambda\left[ f_{0}+\frac{1}{2}f_{0}^{\left(
2,0\right) }\left( x_{2}^{\prime}-x_{20}^{\prime}\right) ^{2}+\frac{1}{2}%
f_{0}^{\left( 0,2\right) }\left( x_{3}^{\prime}-x_{30}^{\prime}\right)
^{2}+\cdots\right] }.
\end{equation}
Expanding the integral up to the second order in $\Lambda$, we obtain the
following expression:%
\begin{equation}
\mathcal{T}(\Lambda)=\sqrt{\frac{2\pi}{\Lambda f_{0}^{\left( 2,0\right) }}}%
\sqrt{\frac{2\pi}{\Lambda f_{0}^{\left( 0,2\right) }}}e^{-\Lambda
f_{0}^{(0,0)}}\left[ u_{0}+\frac{1}{\Lambda}\mathcal{B}(\Lambda^{-1})+\frac{1%
}{\Lambda^{2}}\mathcal{B}\left( \Lambda^{-2}\right) +O\left( \frac{1}{%
\Lambda^{3}}\right) \right]
\end{equation}
where
\begin{align}
\mathcal{B}(\Lambda^{-1}) & =\frac{3}{8}\frac{u_{0}\left( f_{0}^{\left(
2,1\right) }\right) ^{2}}{\left( f_{0}^{\left( 2,0\right) }\right)
^{2}\left( f_{0}^{\left( 0,2\right) }\right) }+\frac{3}{8}\frac {u_{0}\left(
f_{0}^{\left( 1,2\right) }\right) ^{2}}{\left( f_{0}^{\left( 2,0\right)
}\right) \left( f_{0}^{\left( 0,2\right) }\right) ^{2}}+\frac{5}{24}\frac{%
u_{0}\left( f_{0}^{\left( 3,0\right) }\right) ^{2}}{\left( f_{0}^{\left(
2,0\right) }\right) ^{3}}+\frac {5}{24}\frac{u_{0}\left( f_{0}^{\left(
0,3\right) }\right) ^{2}}{\left( f_{0}^{\left( 0,2\right) }\right) ^{3}}
\label{1} \\
& -\frac{1}{4}\frac{u_{0}\left( f_{0}^{\left( 2,2\right) }\right) }{\left(
f_{0}^{\left( 2,0\right) }\right) \left( f_{0}^{\left( 0,2\right) }\right) }-%
\frac{1}{8}\frac{u_{0}\left( f_{0}^{\left( 4,0\right) }\right) }{\left(
f_{0}^{\left( 2,0\right) }\right) ^{2}}-\frac{1}{8}\frac{u_{0}\left(
f_{0}^{\left( 0,4\right) }\right) }{\left( f_{0}^{\left( 0,2\right) }\right)
^{2}}+\frac{1}{4}\frac{u_{0}\left( f_{0}^{\left( 2,1\right) }\right) \left(
f_{0}^{\left( 0,3\right) }\right) }{\left( f_{0}^{\left( 2,0\right) }\right)
\left( f_{0}^{\left( 0,2\right) }\right) ^{2}}\frac{1}{4}\frac{u_{0}\left(
f_{0}^{\left( 1,2\right) }\right) \left( f_{0}^{\left( 3,0\right) }\right) }{%
\left( f_{0}^{\left( 2,0\right) }\right) ^{2}\left( f_{0}^{\left( 0,2\right)
}\right) }  \label{2} \\
& -\frac{1}{2}\frac{u_{0}^{\left( 1,0\right) }\left( f_{0}^{\left(
1,2\right) }\right) }{\left( f_{0}^{\left( 2,0\right) }\right) \left(
f_{0}^{\left( 0,2\right) }\right) }-\frac{1}{2}\frac{u_{0}^{\left(
0,1\right) }\left( f_{0}^{\left( 2,1\right) }\right) }{\left( f_{0}^{\left(
2,0\right) }\right) \left( f_{0}^{\left( 0,2\right) }\right) }-\frac{1}{2}%
\frac{u_{0}^{\left( 1,0\right) }\left( f_{0}^{\left( 3,0\right) }\right) }{%
\left( f_{0}^{\left( 2,0\right) }\right) ^{2}}-\frac{1}{2}\frac{%
u_{0}^{\left( 0,1\right) }\left( f_{0}^{\left( 0,3\right) }\right) }{\left(
f_{0}^{\left( 0,2\right) }\right) ^{2}}  \label{3} \\
& +\frac{1}{2}\frac{u_{0}^{\left( 2,0\right) }}{\left( f_{0}^{\left(
2,0\right) }\right) }+\frac{1}{2}\frac{u_{0}^{\left( 0,2\right) }}{\left(
f_{0}^{\left( 0,2\right) }\right) }.  \label{4}
\end{align}
There are $15$ terms in $\mathcal{B}(\Lambda^{-1})$ above and $151$ terms in
$\mathcal{B}\left( \Lambda^{-2}\right) $ calculated by direct expansion
using Maple, which are consistent with the results we will calculate by hand
in the following.

Indeed, similar to the argument we adopted in Eq.(\ref{sum1}), a typical
term of general higher order $\Lambda^{-M}$ including its coefficient can be
written as%
\begin{align}
\mathcal{B}\left( \Lambda^{-M}\right) & \sim\frac{\left( 2P_{2}\right) !}{%
P_{2}!2^{P_{2}}}\frac{\left( 2P_{3}\right) !}{P_{3}!2^{P_{3}}}\frac {1}{%
m_{2}!m_{3}!}\left[ \underset{n_{2}+n_{3}\geq3}{\prod}\frac{1}{\left(
-n_{2}!n_{3}!\right) ^{V\left( n_{2},n_{3}\right) }V\left(
n_{2},n_{3}\right) !}\right]  \notag \\
& \cdot\frac{u_{0}^{\left( m_{2},m_{3}\right) }\underset{n_{2}+n_{3}\geq3}{%
\prod}\left( f_{0}^{\left( n_{2},n_{3}\right) }\right) ^{V\left(
n_{2},n_{3}\right) }}{\left( f_{0}^{\left( 2,0\right) }\right)
^{P_{2}}\left( f_{0}^{\left( 0,2\right) }\right) ^{P_{3}}}  \label{term1}
\end{align}
where there are $P_{2}$ and $P_{3}$ propogators corresponding to $%
x_{2}^{\prime}$ and $x_{3}^{\prime}$, respectively. In particular, for the
order $\mathcal{B}(\Lambda^{-1})$, Eq.(\ref{term1}) consistently gives all $%
15$ terms in Eq.(\ref{1}) to Eq.(\ref{4}). Here we list some rules regarding
the expansion and the construction of a vacuum diagram corresponds to the
typical term in Eq.(\ref{term1}):

\begin{itemize}
\item $M=$ \# of loops $-$ \# of the connected components $\geq1$,

\item $u_{0}^{\left( m_{2},m_{3}\right) }$ represents a loop with $m_{2}$
external legs corresponding to $f_{0}^{\left( 2,0\right) }$ propagators and,
$m_{3}$ external legs corresponding to $f_{0}^{\left( 0,2\right) }$
propagators, respectively. For the case of $m_{2}=m_{3}=0$, $u_{0}$ will be
treated as a disconnected loop.

\item $f_{0}^{\left( n_{2},n_{3}\right) }$ with $n_{2}+n_{3}\geq3$
represents a vertex with $n_{2}$ legs corresponding to $f_{0}^{\left(
2,0\right) }$ propagators and $n_{3}$ legs corresponding to $f_{0}^{\left(
0,2\right) }$ propagators. $V\left( n_{2},n_{3}\right) $ is the number of $%
f_{0}^{\left( n_{2},n_{3}\right) }$ vertex.

\item $f_{0}^{\left( 2,0\right) }$ and $f_{0}^{\left( 0,2\right) }$ are two
different kinds of propagators.

\item $M$ is the difference between the sum of the numbers of $f_{0}^{\left(
n_{2},n_{3}\right) }$ in the numerator and the sum of numbers of
denominators $f_{0}^{\left( 2,0\right) }$ and $f_{0}^{\left( 0,2\right) }$%
\begin{equation}
P_{2}+P_{3}-\sum_{n_{2}+n_{3}\geq3}V\left( n_{2},n_{3}\right) =M  \label{a}
\end{equation}

\item The number of differentiations with respect to the variables $%
x_{2,3}^{\prime}$ in the numerator equals to the number of differentiations
with respect to the same variables $x_{2,3}^{\prime}$ in the denominator,
respectively%
\begin{align}
m_{2}+\sum_{n_{2}+n_{3}\geq3}n_{2}V\left( n_{2},n_{3}\right) & =2P_{2},
\label{b} \\
m_{3}+\sum_{n_{2}+n_{3}\geq3}n_{3}V\left( n_{2},n_{3}\right) & =2P_{3}.
\label{c}
\end{align}
Eliminating $P_{2}$ and $P_{3}$ from the above constraints, Eq.(\ref{a}),
Eq.(\ref{b}) and Eq.(\ref{c}), we obtain the following equation%
\begin{equation}
\sum_{n_{2}+n_{3}\geq3}\left( n_{2}+n_{3}-2\right) V\left(
n_{2},n_{3}\right) =2M-\left( m_{2}+m_{3}\right) \geq0,  \label{d}
\end{equation}
which is the $5$-point generalization of Eq.(\ref{sol}). We are now ready to
solve Eq.(\ref{d}) order by order. We first define $m=m_{2}+m_{3}$.
\end{itemize}

For the case of $M=1$, the upper bound of $n_{2}+n_{3}$ is $4$ and Eq.(\ref%
{d}) reduces to
\begin{equation}
\sum_{n_{2}+n_{3}=3}^{4}\left( n_{2}+n_{3}-2\right) V\left(
n_{2},n_{3}\right) =2-m\geq0
\end{equation}
or%
\begin{align}
& V\left( 3,0\right) +V\left( 2,1\right) +V\left( 1,2\right) +V\left(
0,3\right)  \notag \\
& +2\left[ V\left( 4,0\right) +V\left( 3,1\right) +V\left( 2,2\right)
+V\left( 1,3\right) +V\left( 0,4\right) \right]  \notag \\
& =2-m.  \label{12}
\end{align}
The $15$ solutions of Eq.(\ref{12}) are listed in the following table%
\begin{equation}
\begin{tabular}{|c|c|c|c|c|c|c|c|c|c|c|c|c|c|c|c|}
\hline
$m_{2}$ & $0$ & $0$ & $0$ & $0$ & $0$ & $0$ & $0$ & $0$ & $0$ & $1$ & $0$ & $%
1$ & $0$ & $2$ & $0$ \\ \hline
$m_{3}$ & $0$ & $0$ & $0$ & $0$ & $0$ & $0$ & $0$ & $0$ & $0$ & $0$ & $1$ & $%
0$ & $1$ & $0$ & $2$ \\ \hline
$V\left( 3,0\right) $ & $2$ & $0$ & $0$ & $0$ & $1$ & $0$ & $0$ & $0$ & $0$
& $1$ & $0$ & $0$ & $0$ & $0$ & $0$ \\ \hline
$V\left( 2,1\right) $ & $0$ & $2$ & $0$ & $0$ & $0$ & $1$ & $0$ & $0$ & $0$
& $0$ & $1$ & $0$ & $0$ & $0$ & $0$ \\ \hline
$V\left( 1,2\right) $ & $0$ & $0$ & $2$ & $0$ & $1$ & $0$ & $0$ & $0$ & $0$
& $0$ & $0$ & $1$ & $0$ & $0$ & $0$ \\ \hline
$V\left( 0,3\right) $ & $0$ & $0$ & $0$ & $2$ & $0$ & $1$ & $0$ & $0$ & $0$
& $0$ & $0$ & $0$ & $1$ & $0$ & $0$ \\ \hline
$V\left( 4,0\right) $ & $0$ & $0$ & $0$ & $0$ & $0$ & $0$ & $1$ & $0$ & $0$
& $0$ & $0$ & $0$ & $0$ & $0$ & $0$ \\ \hline
$V\left( 3,1\right) $ & $0$ & $0$ & $0$ & $0$ & $0$ & $0$ & $0$ & $0$ & $0$
& $0$ & $0$ & $0$ & $0$ & $0$ & $0$ \\ \hline
$V\left( 2,2\right) $ & $0$ & $0$ & $0$ & $0$ & $0$ & $0$ & $0$ & $1$ & $0$
& $0$ & $0$ & $0$ & $0$ & $0$ & $0$ \\ \hline
$V\left( 1,3\right) $ & $0$ & $0$ & $0$ & $0$ & $0$ & $0$ & $0$ & $0$ & $0$
& $0$ & $0$ & $0$ & $0$ & $0$ & $0$ \\ \hline
$V\left( 0,4\right) $ & $0$ & $0$ & $0$ & $0$ & $0$ & $0$ & $0$ & $0$ & $1$
& $0$ & $0$ & $0$ & $0$ & $0$ & $0$ \\ \hline
\end{tabular}
\ \ .
\end{equation}

Note that the first $4$ lines of the table correspond to $4$ terms in Eq.(%
\ref{1}), the $5$th and the $6$th lines correspond to the last $2$ terms of
Eq.(\ref{2}), the $7$th to the $9$th lines correspond to the first $3$ terms
of Eq.(\ref{2}), the $10$th to the $13$th lines correspond to $4$ terms of
Eq.(\ref{3}) and finally the last $2$ line of the table correspond to $2$
terms of Eq.(\ref{4}).

For the case of $M=2$, the upper bound of $n_{2}+n_{3}$ is $6$ and Eq.(\ref%
{d}) reduces to
\begin{equation}
\sum_{n_{2}+n_{3}=3}^{6}\left( n_{2}+n_{3}-2\right) V\left(
n_{2},n_{3}\right) =4-m\geq0,
\end{equation}
which gives%
\begin{equation}
V_{3}+2V_{4}+3V_{5}+4V_{6}=4-m  \label{444}
\end{equation}
where%
\begin{align}
V_{3} & =V\left( 3,0\right) +V\left( 2,1\right) +V\left( 1,2\right) +V\left(
0,3\right) , \\
V_{4} & =\left[ V\left( 4,0\right) +V\left( 3,1\right) +V\left( 2,2\right)
+V\left( 1,3\right) +V\left( 0,4\right) \right] , \\
V_{5} & =V\left( 5,0\right) +V\left( 4,1\right) +V\left( 3,2\right) +V\left(
2,3\right) +V\left( 1,4\right) +V\left( 0,5\right) , \\
V_{6} & =V\left( 6,0\right) +V\left( 5,1\right) +V\left( 4,2\right) +V\left(
3,3\right) +V\left( 2,4\right) +V\left( 1,5\right) +V\left( 0,6\right) .
\end{align}
The $151$ solutions of Eq.(\ref{444}) are listed in the following table%
\begin{equation}
\begin{tabular}{|c|c|c|c|c|c|c|c|c|c|c|c|c|c|c|c|}
\hline
$m_{2}$ & $0$ & $1$ & $0$ & $2$ & $1$ & $0$ & $3$ & $2$ & $1$ & $0$ & $4$ & $%
3$ & $2$ & $1$ & $0$ \\ \hline
$m_{3}$ & $0$ & $0$ & $1$ & $0$ & $1$ & $2$ & $0$ & $1$ & $2$ & $3$ & $0$ & $%
1$ & $2$ & $3$ & $4$ \\ \hline
$\#$ of terms & $70$ & $23$ & $23$ & $9$ & $6$ & $9$ & $2$ & $2$ & $2$ & $2$
& $1$ & $0$ & $1$ & $0$ & $1$ \\ \hline
\end{tabular}%
\end{equation}
where the last line counts the number of solutions for each $(m_{2}$ , $%
m_{3})$. For the case of $m_{2}+m_{3}=0$, for example, one has $(m_{2}$ , $%
m_{3})=(0$ , $0)$ and the $70$ solutions are%
\begin{equation}
\begin{tabular}{|c|c|c|c|c|c|c|c|c|c|c|c|}
\hline
$V_{3}$ & $4$ & $3+1$ & $2+2$ & $2+1+1$ & $1+1+1+1$ & $2$ & $1+1$ & $1$ & $0$
& $0$ & $0$ \\ \hline
$V_{4}$ & $0$ & $0$ & $0$ & $0$ & $0$ & $1$ & $1$ & $0$ & $2$ & $1+1$ & $0$
\\ \hline
$V_{5}$ & $0$ & $0$ & $0$ & $0$ & $0$ & $0$ & $0$ & $1$ & $0$ & $0$ & $0$ \\
\hline
$V_{6}$ & $0$ & $0$ & $0$ & $0$ & $0$ & $0$ & $0$ & $0$ & $0$ & $0$ & $1$ \\
\hline
$\#$ of terms & $4$ & $4$ & $6$ & $4$ & $1$ & $12$ & $14$ & $12$ & $5$ & $4$
& $4$ \\ \hline
\end{tabular}
.
\end{equation}
For the case of $m_{2}+m_{3}=2$, one has $(m_{2}$ , $m_{3})=(1,1)$, $(2,0)$,
$(0,2)$ and there are $6$, $9$, $9$ solutions respectively
\begin{equation}
\ \ \
\begin{tabular}{|l|l|l|l|l|l|l|l|l|}
\hline
$\left( m_{2},m_{3}\right) $ & \multicolumn{2}{|l|}{$\left( 1,1\right) $} &
\multicolumn{3}{|l|}{$\left( 2,0\right) $} & \multicolumn{3}{|l|}{$\left(
0,2\right) $} \\ \hline
$V_{3}$ & $1+1$ & $0$ & $2$ & $1+1$ & $0$ & $2$ & $1+1$ & $0$ \\ \hline
$V_{4}$ & $0$ & $1$ & $0$ & $0$ & $1$ & $0$ & $0$ & $1$ \\ \hline
\# of terms & $4$ & $2$ & $4$ & $2$ & $3$ & $4$ & $2$ & $3$ \\ \hline
\end{tabular}
.
\end{equation}

For the case of $M=3$, there are $1019$ terms in the expansion. In sum, for
the $5$-point $HSSA$ with energy order $M\leq3$, we can count the number of
terms for each $m$%
\begin{equation}
\begin{tabular}{|c|c|c|c|c|c|c|c|c|}
\hline
$m$ & $0$ & $1$ & $2$ & $3$ & $4$ & $5$ & $6$ & total \\ \hline
$M=0$ & $1$ &  &  &  &  &  &  & $1$ \\ \hline
$M=1$ & $9$ & $4$ & $2$ &  &  &  &  & $15$ \\ \hline
$M=2$ & $70$ & $46$ & $24$ & $8$ & $3$ &  &  & $151$ \\ \hline
$M=3$ & $359$ & $312$ & $201$ & $92$ & $39$ & $12$ & $4$ & $1019$ \\ \hline
\end{tabular}
.  \label{terms2}
\end{equation}
Eq.(\ref{terms2}) is the $5$-point generalization of the $4$-point case
calculated in Eq.(\ref{terms}). We expect that there exists some
distribution formula for Eq.(\ref{terms2}) similar to$P(2M-m)$ in Eq.(\ref{N}%
).

\setcounter{equation}{0} \renewcommand{\theequation}{\arabic{section}.%
\arabic{equation}}

\section{Stringy scaling loop expansion of $n$-point Amplitudes}

The most general $n$-points $SSA$ can be written as (after $SL(2,R)$ fixing)%
\begin{equation}
\mathcal{T}(\Lambda)=\int d^{n-3}x_{i}\text{ }u\left( x_{i}\right)
e^{-\Lambda f\left( x_{i}\right) },\text{ }\left( i=2,\cdots,n-2\right) ,
\label{nn}
\end{equation}
where%
\begin{equation}
f=-\underset{i<j}{\sum}\frac{k_{i}\cdot k_{j}}{\Lambda}\ln\left(
x_{j}-x_{i}\right) ,\text{ }\Lambda=-k_{1}\cdot k_{2}.
\end{equation}
For the $n$-point $HSSA$ with $n-1$ tachyons and $1$ high energy state at
mass level $M^{2}=2(N-1)$%
\begin{equation}
\left\vert \left\{ p_{i}\right\} ,2m,2q\right\rangle =\left( \alpha
_{-1}^{T_{1}}\right) ^{N+p_{1}}\left( \alpha_{-1}^{T_{2}}\right)
^{p_{2}}\cdots\left( \alpha_{-1}^{T_{r}}\right) ^{p_{r}}\left(
\alpha_{-1}^{L}\right) ^{2m}\left( \alpha_{-2}^{L}\right) ^{q}\left\vert
0;k\right\rangle
\end{equation}
where $\sum_{i=1}^{r}p_{i}=-2(m+q)$ with\ $r\leq24$, the number of
transverse directions, $u\left( x_{i}\right) $ can be calculated to be%
\begin{equation}
u=\left( k^{T_{1}}\right) ^{N+p_{1}}\left( k^{T_{2}}\right)
^{p_{2}}\cdots\left( k^{T_{r}}\right) ^{p_{r}}\left( k^{L}\right)
^{2m}\left( k^{\prime L}\right) ^{q},  \label{uu}
\end{equation}
where we have defined%
\begin{equation}
k=-\sum_{i\neq2,n}\frac{k_{i}}{x_{i}-x_{2}}.
\end{equation}
We then perform a Taylor expansion on the multi-variables' critical points%
\begin{equation}
\int d^{n-3}x_{i}\text{ }\left[ u\left( x_{i0}\right) +\cdots\right]
e^{-\Lambda\left[ f\left( x_{20}\right) +\frac{1}{2}\underset{i,j}{\sum }%
\frac{\partial^{2}f_{0}}{\partial x_{i}\partial x_{j}}\left(
x_{i}-x_{i0}\right) \left( x_{j}-x_{j0}\right) +\frac{1}{3!}\underset{i,j,k}{%
\sum}\frac{\partial^{3}f_{0}}{\partial x_{i}\partial x_{j}\partial x_{k}}%
\left( x_{i}-x_{i0}\right) \left( x_{j}-x_{j0}\right) \left(
x_{k}-x_{k0}\right) +\cdots\right] }
\end{equation}
where $\left( x_{20},x_{30},\cdots,x_{\left( n-2\right) 0}\right) $ satisfied%
\begin{align}
\frac{\partial f\left( x_{20},x_{30},\cdots,x_{\left( n-2\right) 0}\right) }{%
\partial x_{2}} & =0,  \notag \\
& \vdots  \notag \\
\frac{\partial f\left( x_{20},x_{30},\cdots,x_{\left( n-2\right) 0}\right) }{%
\partial x_{n-2}} & =0.
\end{align}
For the same reason as in the previous $5$-point case, we need to do a
change of variables to eliminate crossing terms and obtain
\begin{align}
& \int d^{n-3}x_{i}^{\prime}\text{ }\left[ u\left(
x_{20}^{\prime},x_{30}^{\prime},\cdots,x_{\left( n-2\right)
0}^{\prime}\right) +\cdots\right]  \notag \\
& \cdot e^{-\Lambda\left[ f\left( x_{20}^{\prime}\right) +\frac{1}{2}%
\underset{i,j}{\sum}\frac{\partial^{2}f\left(
x_{20}^{\prime},x_{30}^{\prime},\cdots,x_{(n-2)0}^{\prime}\right) }{\partial
x_{i}^{\prime }\partial x_{j}^{\prime}}\left(
x_{i}^{\prime}-x_{i0}^{\prime}\right) \left(
x_{j}^{\prime}-x_{j0}^{\prime}\right) +\frac{1}{3!}\underset{i,j,k}{\sum}%
\frac{\partial^{3}f\left( x_{20}^{\prime},x_{30}^{\prime},\cdots,x_{\left(
n-2\right) 0}^{\prime}\right) }{\partial x_{i}^{\prime}\partial
x_{j}^{\prime}\partial x_{k}^{\prime}}\left(
x_{i}^{\prime}-x_{i0}^{\prime}\right) \left( x_{j}^{\prime}-x_{j0}^{\prime
}\right) \left( x_{k}^{\prime}-x_{k0}^{\prime}\right) +\cdots\right] }
\end{align}
where $\left( x_{20}^{\prime},x_{30}^{\prime},\cdots,x_{(n-2)0}^{\prime
}\right) $ satisfied%
\begin{align}
\frac{\partial f\left( x_{20}^{\prime},x_{30}^{\prime},\cdots,x_{\left(
n-2\right) 0}^{\prime}\right) }{\partial x_{2}} & =0,  \notag \\
& \vdots  \notag \\
\frac{\partial f\left( x_{20}^{\prime},x_{30}^{\prime},\cdots,x_{\left(
n-2\right) 0}^{\prime}\right) }{\partial x_{n-2}} & =0.
\end{align}

We define the coefficients in the Taylor expansion of $f$ and $u$ at $\left(
x_{20}^{\prime},x_{30}^{\prime},\cdots,x_{\left( n-2\right) 0}^{\prime
}\right) $ as follows%
\begin{align}
u\left( x_{20}^{\prime},x_{30}^{\prime},\cdots,x_{\left( n-2\right)
0}^{\prime}\right) & =u_{0}, \\
\frac{\partial^{m_{2}+\cdots+m_{n-2}}u\left( x_{20}^{\prime},x_{30}^{\prime
},\cdots,x_{\left( n-2\right) 0}^{\prime}\right) }{\partial\left(
x_{2}^{\prime}\right) ^{m_{2}}\cdots\partial\left( x_{n-2}^{\prime}\right)
^{m_{n-2}}} & =u_{0}^{\left( m_{2},\cdots,m_{n-2}\right) }=u_{0}^{\left\{
m_{i}\right\} }, \\
\frac{\partial^{n_{2}+\cdots+n_{n-2}}f\left( x_{20}^{\prime},x_{30}^{\prime
},\cdots,x_{\left( n-2\right) 0}^{\prime}\right) }{\partial\left(
x_{2}^{\prime}\right) ^{n_{2}}\cdots\partial\left( x_{n-2}^{\prime}\right)
^{n_{n-2}}} & =f_{0}^{\left( n_{2},\cdots,n_{n-2}\right) }=f_{0}^{\left\{
n_{i}\right\} }.
\end{align}

We can then simplify the integral into the following form
\begin{align}
&  \int d^{n-3}x_{i}^{\prime}\text{ }\left[  u_{0}+\cdots\right]
e^{-\Lambda\left[  f_{0}+\frac{1}{2}\sum\limits_{i=2}^{n-2}f_{0}^{\left\{
n_{i}\right\}  }\left(  x_{i}^{\prime}-x_{i0}^{\prime}\right)  ^{2}+\frac
{1}{3!}\sum\limits_{n_{2}+\cdots+n_{n-2}=3}\left(  f_{0}^{\left\{
n_{i}\right\}  }%
%TCIMACRO{\dprod \limits_{i=2}^{n-2}}%
%BeginExpansion
{\displaystyle\prod\limits_{i=2}^{n-2}}
%EndExpansion
\left(  x_{i}^{\prime}-x_{i0}^{\prime}\right)  ^{n_{i}}\right)  +\cdots
\right]  }\nonumber\\
&  =%
%TCIMACRO{\dprod \limits_{j=2}^{n-2}}%
%BeginExpansion
{\displaystyle\prod\limits_{j=2}^{n-2}}
%EndExpansion
\sqrt{\frac{2\pi}{\Lambda\partial_{j}^{2}f_{0}}}e^{-\Lambda f_{0}^{\left\{
0\right\}  }}\left[  u_{0}+\frac{1}{\Lambda}\mathcal{C}(\Lambda^{-1})+\frac
{1}{\Lambda^{2}}\mathcal{C}\left(  \Lambda^{-2}\right)  +O\left(  \frac
{1}{\Lambda^{3}}\right)  \right]  .
\end{align}
After performing the integrations, a typical term in order $\Lambda^{-M}$ of
the above equation can be written as%
\begin{equation}
\mathcal{C}\left(  \Lambda^{-M}\right)  \sim\left[
%TCIMACRO{\dprod \limits_{j=2}^{n-2}}%
%BeginExpansion
{\displaystyle\prod\limits_{j=2}^{n-2}}
%EndExpansion
\frac{\left(  2P_{j}\right)  !}{P_{j}!2^{P_{j}}}\frac{1}{m_{j}!}\right]
\cdot\left[  \underset{\Sigma n_{i}\geq3}{\prod}\frac{1}{V\left(  \left\{
n_{i}\right\}  \right)  !}\left(  -%
%TCIMACRO{\dprod \limits_{j=2}^{n-2}}%
%BeginExpansion
{\displaystyle\prod\limits_{j=2}^{n-2}}
%EndExpansion
n_{j}!\right)  ^{-V\left(  \left\{  n_{i}\right\}  \right)  }\right]
\cdot\frac{u_{0}^{\left\{  m_{i}\right\}  }\underset{\Sigma n_{i}\geq3}{\prod
}\left(  f_{0}^{\left\{  n_{i}\right\}  }\right)  ^{V\left(  \left\{
n_{i}\right\}  \right)  }}{%
%TCIMACRO{\dprod \limits_{j=2}^{n-2}}%
%BeginExpansion
{\displaystyle\prod\limits_{j=2}^{n-2}}
%EndExpansion
\left(  \partial_{j}^{2}f_{0}\right)  ^{P_{j}}} \label{coeff}%
\end{equation}
where $V\left( \left\{ n_{i}\right\} \right) $ is the number of $%
f_{0}^{\left\{ n_{i}\right\} }$ vertex and there are $P_{j}$ propogators
corresponding to $x_{j}^{\prime}$, $j=2,3\cdots,n-2$. Similar rules after
Eq.(\ref{term1}) can be easily set up. Moreover, for the $n$-point case, Eq.(%
\ref{a}) is now replaced by
\begin{equation}
\sum_{j=2}^{n-2}P_{j}-\sum_{n_{2}+\cdots+n_{n-2}\geq3}V\left( \left\{
n_{i}\right\} \right) =M,  \label{aa}
\end{equation}
and Eq.(\ref{b}) and Eq.(\ref{c}) are replaced by

\begin{equation}
m_{j}+\sum_{n_{2}+\cdots+n_{n-2}\geq3}n_{j}V\left( \left\{ n_{i}\right\}
\right) =2P_{j}.\left( j=2,\cdots,n-2\right) .  \label{bb}
\end{equation}
Finally, eliminating $P_{j}$ from the above constraints, Eq.(\ref{aa}) and
Eq.(\ref{bb}), we obtain the following equation

\begin{equation}
\left( \sum_{n_{2}+\cdots+n_{n-2}\geq3}n_{2}+n_{3}+\cdots+n_{n-2}-2\right)
V\left( \left\{ n_{i}\right\} \right) =2M-\sum_{j=2}^{n-2}m_{j}\geq0.
\label{oo}
\end{equation}
which is the $n$-point generalization of Eq.(\ref{d}) and Eq.(\ref{sol}).
One can now solve Eq.(\ref{oo}) order by order as we did previously for the $%
4$-point and $5$-point cases.

\setcounter{equation}{0} \renewcommand{\theequation}{\arabic{section}.%
\arabic{equation}}

\section{Vacuum diagram representation of HSSA}

In this section, similar to the Feynman diagram representation in field
theory, we give a vacuum diagram representation for stringy scaling loop
expansion of $HSSA$. We will see that in general for each term of the
expansion, there can be many diagrams correspond to it. In particular, we
will sum over the inverse \textit{symmetry factors} of all diagrams of the
term to consistently match with the coefficient of the term.

We begin with the $4$-point $HSSA$ with order $M=1$, namely Eq.(\ref{1111})
to Eq.(\ref{4444}). The corresponding diagrams are
\begin{align}
\mathcal{A}_{1}(\Lambda^{-1}) & =-\frac{u_{0}f_{0}^{(4)}}{%
8(f_{0}^{\prime\prime})^{2}}=-\frac{1}{2^{3}}\raisebox{-1ex}{%
\includegraphics[scale=0.5]{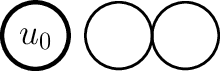}}, \\
\mathcal{A}_{2}(\Lambda^{-1}) & =\frac{5u_{0}(f_{0}^{(3)})^{2}}{%
24(f_{0}^{\prime\prime})^{3}}=\frac{1}{2^{3}}\raisebox{-1ex}{%
\includegraphics[scale=0.5]{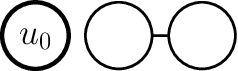}}+\frac {1}{2\cdot3!}%
\raisebox{-1ex}{\includegraphics[scale=0.5]{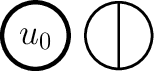}},  \label{2D}
\\
\mathcal{A}_{3}(\Lambda^{-1}) & =-\frac{u_{0}^{\prime}f_{0}^{(3)}}{%
2(f_{0}^{\prime\prime})^{2}}=-\frac{1}{2}\raisebox{-1ex}{%
\includegraphics[scale=0.5]{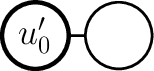}},  \label{3D} \\
\mathcal{A}_{4}(\Lambda^{-1}) & =\frac{u_{0}^{\prime\prime}}{%
2f_{0}^{\prime\prime}}=\frac{1}{2}\raisebox{-1ex}{%
\includegraphics[scale=0.5]{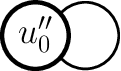}}.
\end{align}

We see that there are two diagrams corresponding to one term in Eq.(\ref{2D}%
). We will see that there will be even more diagrams corresponding to one
term in the higher order expansion as will see next.

We are now ready to use the rules listed in Eq.(\ref{FR1}) to draw the
vacuum diagrams. For Eq.(\ref{2D}) as the first example, one wants to draw
all vacuum diagrams with $3$ propagators $f_{0}^{\prime\prime}$ , $2$ $3$%
-point vertex $f_{0}^{(3)}$ and a disconnected loop corresponding to $u_{0}$%
. There are two diagrams for this term and $M$ for each diagram is $M=3-2=1$%
. Moreover, the sum of the inverse symmetry factor%
\begin{equation}
\frac{1}{2^{3}}+\frac{1}{2\cdot3!}=\frac{5}{24}
\end{equation}
is consistent with the general formula calculated in Eq.(\ref{sum1}).
Indeed, for $P=3$, $m=0$, $n=3$ and $V(3)=2$, the coefficient calculated in
Eq.(\ref{sum1}) is
\begin{equation}
\frac{\left( 2P\right) !}{P!2^{P}}\frac{1}{m!}\left[ \underset{n\geq 3}{\prod%
}\frac{1}{\left( -n!\right) ^{V\left( n\right) }V\left( n\right) !}\right] =%
\frac{5}{24}.
\end{equation}
For Eq.(\ref{3D}) as the second example, one wants to draw all vacuum
diagrams with $2$ propagators $f_{0}^{\prime\prime}$ , $1$ $3$-point vertex $%
f_{0}^{(3)}$ and a tadpole corresponding to $u_{0}^{\prime}$. There is only
one diagram for this term and the value of its $M$ is $M=2-1=1$.

We next consider the $4$-point $HSSA$ with order $M=2$, namely Eq.(\ref{2222}%
) to Eq.(\ref{3333}). The diagram representations including the inverse
symmetry factors for each term are
\begin{align}
\mathcal{A}_{1}(\Lambda^{-2}) & =-\frac{u_{0}f_{0}^{(6)}}{%
48(f_{0}^{\prime\prime})^{3}}=-\frac{1}{3!\cdot2^{3}}\raisebox{-3ex}{%
\includegraphics[scale=0.5]{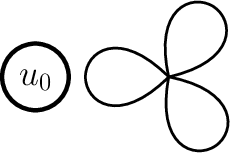}}, \\
\mathcal{A}_{2}(\Lambda^{-2}) & =\frac{7u_{0}f_{0}^{(3)}f_{0}^{(5)}}{%
48(f_{0}^{\prime\prime})^{4}}=\frac{1}{3!\cdot2!}\raisebox{-1ex}{%
\includegraphics[scale=0.5]{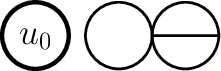}}+\frac {1}{2\cdot2^{3}}%
\raisebox{-3ex}{\includegraphics[scale=0.5]{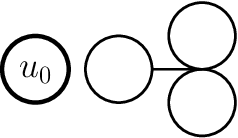}}, \\
\mathcal{A}_{3}(\Lambda^{-2}) & =\frac{35u_{0}(f_{0}^{(4)})^{2}}{%
384(f_{0}^{\prime\prime})^{4}}=\frac{1}{2\cdot4!}\raisebox{-1ex}{%
\includegraphics[scale=0.5]{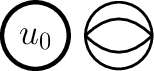}}+\frac{1}{2^{4}}%
\raisebox{-1ex}{\includegraphics[scale=0.5]{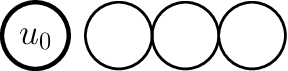}}+\frac {1}{%
2\cdot8^{2}}\raisebox{-1ex}{\includegraphics[scale=0.5]{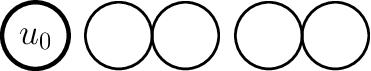}},
\\
\mathcal{A}_{4}(\Lambda^{-2}) & =-\frac{35u_{0}f_{0}^{(4)}(f_{0}^{(3)})^{2}}{%
64(f_{0}^{\prime\prime})^{5}}=-\frac{1}{8}\raisebox{-1ex}{%
\includegraphics[scale=0.5]{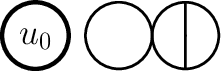}}-\frac {1}{2\cdot3!}%
\raisebox{-1ex}{\includegraphics[scale=0.5]{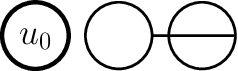}}-\frac{1}{8}%
\raisebox{-1ex}{\includegraphics[scale=0.5]{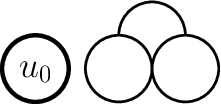}}  \notag \\
& -\frac{1}{8}\raisebox{-1ex}{\includegraphics[scale=0.5]{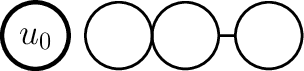}}%
-\frac{1}{16}\raisebox{-3ex}{\includegraphics[scale=0.5]{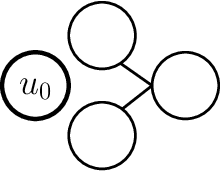}}-%
\frac {1}{8\cdot2\cdot3!}\raisebox{-1ex}{%
\includegraphics[scale=0.5]{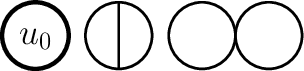}}-\frac {1}{8\cdot8}%
\raisebox{-1ex}{\includegraphics[scale=0.5]{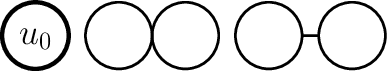}}, \\
\mathcal{A}_{5}(\Lambda^{-2}) & =\frac{385u_{0}(f_{0}^{(3)})^{4}}{%
1152(f_{0}^{\prime\prime})^{6}}=\frac{1}{4!}\raisebox{-1ex}{%
\includegraphics[scale=0.5]{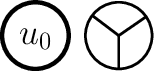}}+\frac{1}{16}%
\raisebox{-1ex}{\includegraphics[scale=0.5]{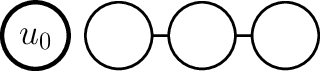}}+\frac {1}{%
3!\cdot2^{3}}\raisebox{-3ex}{\includegraphics[scale=0.5]{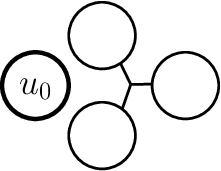}}
\notag \\
& +\frac{1}{2^{3}}\raisebox{-1ex}{%
\includegraphics[scale=0.5]{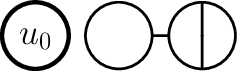}}+\frac{1}{2^{4}}%
\raisebox{-1ex}{\includegraphics[scale=0.5]{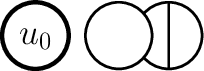}}+\frac {1}{%
2\cdot2^{6}}\raisebox{-1ex}{\includegraphics[scale=0.5]{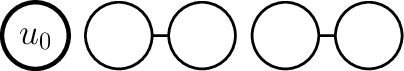}}
\notag \\
& +\frac{1}{2^{3}\cdot3!\cdot2}\raisebox{-1ex}{%
\includegraphics[scale=0.5]{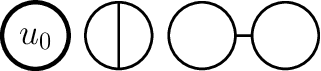}}+\frac{1}{2^{2}\cdot\left(
3!\right) ^{2}\cdot2}\raisebox{-1ex}{%
\includegraphics[scale=0.5]{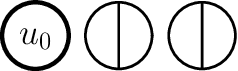}},  \label{59} \\
\mathcal{A}_{6}(\Lambda^{-2}) & =-\frac{u_{0}^{^{\prime}}f_{0}^{(5)}}{%
8(f_{0}^{\prime\prime})^{3}}=-\frac{1}{2^{3}}\raisebox{-3ex}{%
\includegraphics[scale=0.5]{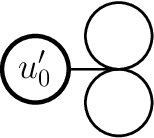}}, \\
\mathcal{A}_{7}(\Lambda^{-2}) & =\frac{35u_{0}^{^{%
\prime}}f_{0}^{(3)}f_{0}^{(4)}}{48(f_{0}^{\prime\prime})^{4}}=\frac{1}{2^{2}}%
\raisebox{-1ex}{\includegraphics[scale=0.5]{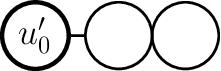}}+\frac{1}{3!}%
\raisebox{-1ex}{\includegraphics[scale=0.5]{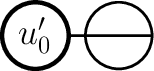}}+\frac {1}{%
2^{2}}\raisebox{-1ex}{\includegraphics[scale=0.5]{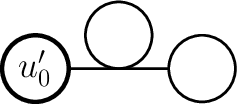}}+\frac{1%
}{2^{4}}\raisebox{-1ex}{\includegraphics[scale=0.5]{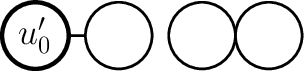}},
\label{61} \\
\mathcal{A}_{8}(\Lambda^{-2}) & =-\frac{35u_{0}^{^{\prime}}(f_{0}^{(3)})^{2}%
}{48(f_{0}^{\prime\prime})^{5}}=-\frac{1}{2^{2}}\raisebox{-1ex}{%
\includegraphics[scale=0.5]{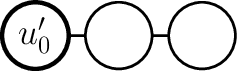}}-\frac{1}{2^{2}}%
\raisebox{-1ex}{\includegraphics[scale=0.5]{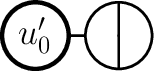}}-\frac {1}{8}%
\raisebox{-3ex}{\includegraphics[scale=0.5]{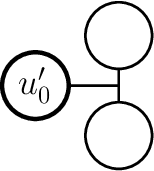}}  \notag \\
& -\frac{1}{2\cdot2^{3}}\raisebox{-1ex}{%
\includegraphics[scale=0.5]{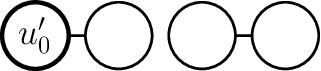}}-\frac {1}{2\cdot2\cdot3!}%
\raisebox{-1ex}{\includegraphics[scale=0.5]{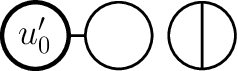}}, \\
\mathcal{A}_{9}(\Lambda^{-2}) & =-\frac{5u_{0}^{^{\prime\prime}}f_{0}^{(4)}}{%
16(f_{0}^{\prime\prime})^{3}}=-\frac{1}{4}\raisebox{-1ex}{%
\includegraphics[scale=0.5]{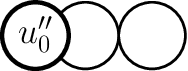}}-\frac {1}{2\cdot2^{3}}%
\raisebox{-1ex}{\includegraphics[scale=0.5]{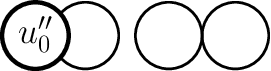}}, \\
\mathcal{A}_{10}(\Lambda^{-2}) & =\frac{35u_{0}^{\prime%
\prime}(f_{0}^{(3)})^{2}}{48(f_{0}^{\prime\prime})^{4}}=\frac{1}{2^{2}}%
\raisebox{-1ex}{\includegraphics[scale=0.5]{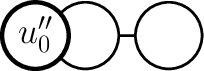}}+\frac {1}{%
2^{2}\cdot3!}\raisebox{-1ex}{\includegraphics[scale=0.5]{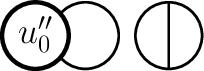}}%
+\frac {1}{2\cdot2^{2}}\raisebox{-1ex}{%
\includegraphics[scale=0.5]{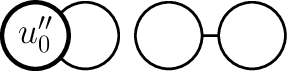}}  \notag \\
& +\frac{1}{4}\raisebox{-1ex}{%
\includegraphics[scale=0.5]{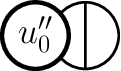}}+\frac{1}{8}%
\raisebox{-1ex}{\includegraphics[scale=0.5]{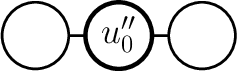}}, \\
\mathcal{A}_{11}(\Lambda^{-2}) & =-\frac{5u_{0}^{(3)}f_{0}^{(3)}}{%
12(f_{0}^{\prime\prime})^{3}}=-\frac{1}{3!}\raisebox{-1ex}{%
\includegraphics[scale=0.5]{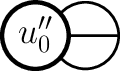}}-\frac {1}{2^{2}}%
\raisebox{-1ex}{\includegraphics[scale=0.5]{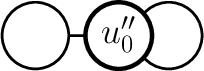}}, \\
\mathcal{A}_{12}(\Lambda^{-2}) & =\frac{u_{0}^{(4)}}{8(f_{0}^{\prime\prime
})^{2}}=\frac{1}{2^{3}}\raisebox{-1ex}{%
\includegraphics[scale=0.5]{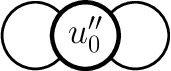}}.
\end{align}

It is important to note that the coefficient of each term in $\mathcal{A}%
_{j}(\Lambda^{-1})$ and $\mathcal{A}_{k}(\Lambda^{-2})$ matches with the sum
of the inverse symmetry factors of all diagrams corresponding to the term.
For the example of the term $\mathcal{A}_{5}(\Lambda^{-2})$ in Eq.(\ref{59}%
), there are $8$ diagrams corresponding to it. The sum of the inverse
symmetry factors \cite{Peskin} gives%
\begin{equation}
\frac{1}{4!}+\frac{1}{16}+\frac{1}{3!\cdot2^{3}}+\frac{1}{2^{3}}+\frac {1}{%
2^{4}}+\frac{1}{2\cdot2^{6}}+\frac{1}{2^{3}\cdot3!\cdot2}+\frac{1}{%
2^{2}\cdot\left( 3!\right) ^{2}\cdot2}=\frac{385}{1152}=\frac{\left(
2P\right) !}{P!2^{P}}\frac{1}{m!}\underset{n\geq3}{\prod}\left( \frac{\left(
\frac{-1}{n!}\right) ^{V\left( n\right) }}{V\left( n\right) !}\right) ,
\label{000}
\end{equation}
which is consistent with Eq.(\ref{sum1}) for $P=6$, $m=0$, $n=3$ and $V(3)=4$%
. The result of this coefficient is also consistent with Eq.(\ref{main}).
Note that, to the order $\mathcal{A}_{k}(\Lambda^{-2})$, there are $3,4,5$
and $6$-point vertices in the diagrams which are much more than those in the
case of usual quantum field theory.

We now use the rules listed in Eq.(\ref{FR1}) to draw all the vacuum
diagrams corresponding to the term in Eq.(\ref{59}). One wants to draw all
vacuum diagrams with $6$ propagators $f_{0}^{\prime\prime}$ , $4$ $3$-point
vertex $f_{0}^{(3)}$ and a disconnected loop corresponding to $u_{0}$. There
are $8$ diagrams for this term and $M$ for each diagram is $M=4-2=5-3=2$%
.\bigskip

As the second example, we wants to draw all the vacuum diagrams
corresponding to the term in Eq.(\ref{61}). One has to draw all vacuum
diagrams with $4$ propagators $f_{0}^{\prime\prime}$ , $1$ $3$-point vertex $%
f_{0}^{(3)}$, $1$ $4$-point vertex $f_{0}^{(4)}$and a tadpole corresponding
to $u_{0}^{\prime}$. There are $4$ diagrams for this term and $M$ for each
diagram is $M=4-2=3-1=2$.\bigskip\ The sum of the inverse symmetry factors
gives%
\begin{equation}
\frac{1}{4!}+\frac{1}{16}+\frac{1}{3!\cdot2^{3}}+\frac{1}{2^{3}}+\frac {1}{%
2^{4}}+\frac{1}{2\cdot2^{6}}+\frac{1}{2^{3}\cdot3!\cdot2}+\frac{1}{%
2^{2}\cdot\left( 3!\right) ^{2}\cdot2}=\frac{35}{48}=\frac{\left( 2P\right) !%
}{P!2^{P}}\frac{1}{m!}\underset{n\geq3}{\prod}\left( \frac{\left( \frac{-1}{%
n!}\right) ^{V\left( n\right) }}{V\left( n\right) !}\right) ,
\end{equation}
which is consistent with Eq.(\ref{sum1}) for $P=4$, $m=1$, $n=3$, $V(4)=1$
and $V(3)=1$.

There are $30$ terms of $4$-point $HSSA$ $\mathcal{A}_{j}(\Lambda^{-3})$
with order $M=3$. The corresponding diagrams can be similarly written down.

The $5$-point $HSSA$ with order $M=1$ are%
\begin{align}
\mathcal{B}_{1}(\Lambda^{-1}) & =\frac{3}{8}\frac{u_{0}\left( f_{0}^{\left(
2,1\right) }\right) ^{2}}{\left( f_{0}^{\left( 2,0\right) }\right)
^{2}\left( f_{0}^{\left( 0,2\right) }\right) }=\frac{1}{2^{2}}%
\raisebox{-1ex}{\includegraphics[scale=0.5]{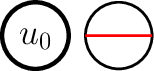}}+\frac {1}{%
2^{3}}\raisebox{-1ex}{\includegraphics[scale=0.5]{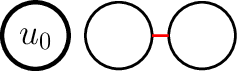}}, \\
\mathcal{B}_{2}(\Lambda^{-1}) & =\frac{3}{8}\frac{u_{0}\left( f_{0}^{\left(
1,2\right) }\right) ^{2}}{\left( f_{0}^{\left( 2,0\right) }\right) \left(
f_{0}^{\left( 0,2\right) }\right) ^{2}}=\frac{1}{2^{2}}\raisebox{-1ex}{%
\includegraphics[scale=0.5]{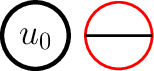}}+\frac {1}{2^{3}}%
\raisebox{-1ex}{\includegraphics[scale=0.5]{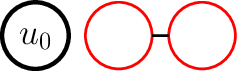}},  \label{pp}
\\
\mathcal{B}_{3}(\Lambda^{-1}) & =\frac{5}{24}\frac{u_{0}\left( f_{0}^{\left(
3,0\right) }\right) ^{2}}{\left( f_{0}^{\left( 2,0\right) }\right) ^{3}}=%
\frac{1}{2\cdot3!}\raisebox{-1ex}{%
\includegraphics[scale=0.5]{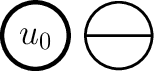}}+\frac {1}{2^{3}}%
\raisebox{-1ex}{\includegraphics[scale=0.5]{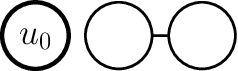}}, \\
\mathcal{B}_{4}(\Lambda^{-1}) & =\frac{5}{24}\frac{u_{0}\left( f_{0}^{\left(
0,3\right) }\right) ^{2}}{\left( f_{0}^{\left( 0,2\right) }\right) ^{3}}=%
\frac{1}{2\cdot3!}\raisebox{-1ex}{%
\includegraphics[scale=0.5]{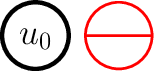}}+\frac {1}{2^{3}}%
\raisebox{-1ex}{\includegraphics[scale=0.5]{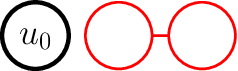}}, \\
\mathcal{B}_{5}(\Lambda^{-1}) & =-\frac{1}{4}\frac{u_{0}\left( f_{0}^{\left(
2,2\right) }\right) }{\left( f_{0}^{\left( 2,0\right) }\right) \left(
f_{0}^{\left( 0,2\right) }\right) }=-\frac{1}{2^{2}}\raisebox{-1ex}{%
\includegraphics[scale=0.5]{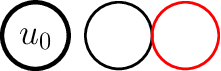}}, \\
\mathcal{B}_{6}(\Lambda^{-1}) & =-\frac{1}{8}\frac{u_{0}\left( f_{0}^{\left(
4,0\right) }\right) }{\left( f_{0}^{\left( 2,0\right) }\right) ^{2}}=-\frac{1%
}{2\cdot2!}\raisebox{-1ex}{\includegraphics[scale=0.5]{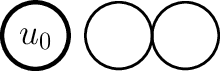}}, \\
\mathcal{B}_{7}(\Lambda^{-1}) & =-\frac{1}{8}\frac{u_{0}\left( f_{0}^{\left(
0,4\right) }\right) }{\left( f_{0}^{\left( 0,2\right) }\right) ^{2}}=-\frac{1%
}{2\cdot2!}\raisebox{-1ex}{\includegraphics[scale=0.5]{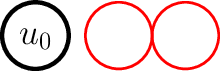}}, \\
\mathcal{B}_{8}(\Lambda^{-1}) & =\frac{1}{4}\frac{u_{0}\left( f_{0}^{\left(
2,1\right) }\right) \left( f_{0}^{\left( 0,3\right) }\right) }{\left(
f_{0}^{\left( 2,0\right) }\right) \left( f_{0}^{\left( 0,2\right) }\right)
^{2}}=\frac{1}{2^{2}}\raisebox{-1ex}{%
\includegraphics[scale=0.5]{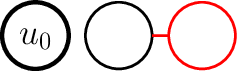}}, \\
\mathcal{B}_{9}(\Lambda^{-1}) & =\frac{1}{4}\frac{u_{0}\left( f_{0}^{\left(
1,2\right) }\right) \left( f_{0}^{\left( 3,0\right) }\right) }{\left(
f_{0}^{\left( 2,0\right) }\right) ^{2}\left( f_{0}^{\left( 0,2\right)
}\right) }=\frac{1}{2^{2}}\raisebox{-1ex}{%
\includegraphics[scale=0.5]{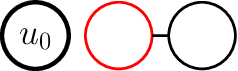}}, \\
\mathcal{B}_{10}(\Lambda^{-1}) & =-\frac{1}{2}\frac{u_{0}^{\left( 1,0\right)
}\left( f_{0}^{\left( 1,2\right) }\right) }{\left( f_{0}^{\left( 2,0\right)
}\right) \left( f_{0}^{\left( 0,2\right) }\right) }=-\frac{1}{2}%
\raisebox{-1ex}{\includegraphics[scale=0.5]{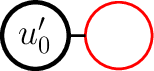}}, \\
\mathcal{B}_{11}(\Lambda^{-1}) & =-\frac{1}{2}\frac{u_{0}^{\left( 0,1\right)
}\left( f_{0}^{\left( 2,1\right) }\right) }{\left( f_{0}^{\left( 2,0\right)
}\right) \left( f_{0}^{\left( 0,2\right) }\right) }=-\frac{1}{2}%
\raisebox{-1ex}{\includegraphics[scale=0.5]{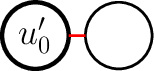}}, \\
\mathcal{B}_{12}(\Lambda^{-1}) & =-\frac{1}{2}\frac{u_{0}^{\left( 1,0\right)
}\left( f_{0}^{\left( 3,0\right) }\right) }{\left( f_{0}^{\left( 2,0\right)
}\right) ^{2}}=-\frac{1}{2}\raisebox{-1ex}{%
\includegraphics[scale=0.5]{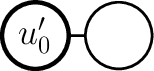}}, \\
\mathcal{B}_{13}(\Lambda^{-1}) & =-\frac{1}{2}\frac{u_{0}^{\left( 0,1\right)
}\left( f_{0}^{\left( 0,3\right) }\right) }{\left( f_{0}^{\left( 0,2\right)
}\right) ^{2}}=-\frac{1}{2}\raisebox{-1ex}{%
\includegraphics[scale=0.5]{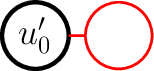}}, \\
\mathcal{B}_{14}(\Lambda^{-1}) & =\frac{1}{2}\frac{u_{0}^{\left( 2,0\right) }%
}{\left( f_{0}^{\left( 2,0\right) }\right) }=\frac{1}{2}\raisebox{-1ex}{%
\includegraphics[scale=0.5]{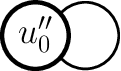}}, \\
\mathcal{B}_{15}(\Lambda^{-1}) & =\frac{1}{2}\frac{u_{0}^{\left( 0,2\right) }%
}{\left( f_{0}^{\left( 0,2\right) }\right) }=\frac{1}{2}\raisebox{-1ex}{%
\includegraphics[scale=0.5]{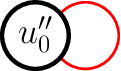}}
\end{align}
where black lines represent the propagators corresponding to $f_{0}^{\left(
2,0\right) }$, and red lines represent the propagators corresponding to $%
f_{0}^{\left( 0,2\right) }$. The coefficient of each term in $\mathcal{B}%
_{j}(\Lambda^{-1})$ also matches with the sum of the inverse symmetry
factors of all diagrams corresponding to the term.

\bigskip As an example, we use the rules listed before Eq.(\ref{a}) to draw
all the vacuum diagrams corresponding to the term in Eq.(\ref{pp}). One
wants to draw all vacuum diagrams with $1$ propagator $f_{0}^{\left(
2,0\right) }$ , $2$ propagators $f_{0}^{\left( 0,2\right) }$, $2$ $3$-point
vertex $4$ $3$-point vertex $f_{0}^{\left( 0,2\right) }$ and a disconnected
loop corresponding to $u_{0}$. There are $2$ diagrams for this term and $M$
for each diagram is $M=3-2=1$.\bigskip\ The sum of the inverse symmetry
factors gives%
\begin{equation}
\frac{1}{2^{2}}+\frac{1}{2^{3}}=\frac{3}{8}=\frac{\left( 2P_{2}\right) !}{%
P_{2}!2^{P_{2}}}\frac{\left( 2P_{3}\right) !}{P_{3}!2^{P_{3}}}\frac {1}{%
m_{2}!m_{3}!}\left[ \underset{n_{2}+n_{3}\geq3}{\prod}\frac{1}{\left(
-n_{2}!n_{3}!\right) ^{V\left( n_{2},n_{3}\right) }V\left(
n_{2},n_{3}\right) !}\right] ,
\end{equation}
which is consistent with Eq.(\ref{term1}) for $P_{2}=1$, $P_{3}=2$, $%
m_{2}=m_{3}=0$, $n_{2}=1$, $n_{3}=2$ and $V(1,2)=2$.

There are $151$ terms of $5$-point $HSSA$ $\mathcal{B}_{j}(\Lambda^{-2})$
with order $M=2$. The corresponding diagrams can be similarly written down.

\setcounter{equation}{0} \renewcommand{\theequation}{\arabic{section}.%
\arabic{equation}}

\section{Stringy scaling violation}

In this section, we apply the stringy scaling loop expansion developed in
the previous sections to calculate the $HSSA$. We begin with the $4$-point $%
HSSA$. For this case it has been known that all leading order $HSSA$ at each
fixed mass level share the same functional form and is independent of the
scattering angle $\phi$. The ratios among $4$-point $HSSA$ at a fixed mass
level $N$ was calculated to be \cite{ChanLee,ChanLee2,CHLTY2,CHLTY1}

\begin{equation}
\frac{\mathcal{T}^{\left( N,2m,q\right) }}{\mathcal{T}^{\left( N,0,0\right) }%
}=\frac{\left( 2m\right) !}{m!}\left( \frac{-1}{2M_{2}}\right) ^{2m+q}.\text{%
(\textbf{independent of }}\phi\text{ !)}  \label{22}
\end{equation}
In Eq.(\ref{22}) $\mathcal{T}^{\left( N,2m,q\right) }$ is the $4$-point $%
HSSA $ of any string vertex $V_{j}$ with $j=1,3,4$ and $V_{2}$ is the high
energy state in Eq.(\ref{111}); and $\mathcal{T}^{\left( N,0,0\right) }$ is
the $4$-point $HSSA$ of any string vertex $V_{j}$ with $j=1,3,4$, and $V_{2}$
is the leading Regge trajectory string state at mass level $N$. Note that in
Eq.(\ref{22}) we have omitted the tensor indice of $V_{j}$ with $j=1,3,4$
and keep only those of $V_{2}$ in $\mathcal{T}^{\left( N,2m,q\right) }$. It
is important to note that to calculate the \textit{nontrivial leading order}
amplitude $\mathcal{T}^{\left( N,2m,q\right) }$, one needs to calculate the $%
HSSA$ up to the order $\frac{1}{\Lambda^{m}}$. As an example, for the case
of $N=3$ in Eq.(\ref{111}), Eq.(\ref{333}) leads to

\begin{align}
\left( \alpha_{-1}^{T}\right) ^{3}\left\vert 0;k\right\rangle ,\left(
m,q\right) & =\left( 0,0\right) ,\mathcal{T}^{\left( 3,0,0\right) }\sim\frac{%
\frac{1}{4}\sqrt{2}\Lambda^{\frac{3}{2}}\left( -1+\tau\right) ^{\frac{9}{2}}%
}{\tau^{\frac{3}{2}}}  \label{a1} \\
+ & \frac{\frac{1}{48}\sqrt{2}\sqrt{\Lambda}\left( -1+\tau\right) ^{\frac{3}{%
2}}\left( \tau^{4}-27\tau^{3}+88\tau^{2}-99\tau+37\right) }{-\tau^{\frac{5}{2%
}}},  \label{a2} \\
\left( \alpha_{-1}^{T}\right) \left( \alpha_{-2}^{L}\right) \left\vert
0;k\right\rangle ,\left( m,q\right) & =\left( 0,1\right) ,\mathcal{T}%
^{\left( 3,0,1\right) }\sim\frac{\frac{1}{2}\sqrt{2}\Lambda^{\frac{3}{2}%
}\left( -1+\tau\right) ^{\frac{9}{2}}}{M\tau^{\frac{3}{2}}}  \label{a3} \\
+ & \frac{\frac{1}{24}\sqrt{2}\sqrt{\Lambda}\left( -1+\tau\right) ^{\frac{3}{%
2}}\left( 13\tau^{4}-15\tau^{3}+28\tau^{2}-63\tau+37\right) }{\tau^{\frac{5}{%
2}}M},  \label{a4} \\
\left( \alpha_{-1}^{T}\right) \left( \alpha_{-1}^{L}\right) ^{2}\left\vert
0;k\right\rangle ,\left( m,q\right) & =\left( 1,0\right) ,\mathcal{T}%
^{\left( 3,2,0\right) }\sim\frac{\frac{1}{2}\sqrt{2}\Lambda^{\frac{3}{2}%
}\left( -1+\tau\right) ^{\frac{9}{2}}}{M\tau^{\frac{3}{2}}}  \label{a5} \\
& +\frac{\frac{1}{24}\sqrt{2}\sqrt{\Lambda}\left( -1+\tau\right) ^{\frac {3}{%
2}}\left( 13\tau^{4}-15\tau^{3}+52\tau^{2}-111\tau+61\right) }{-\tau^{\frac{5%
}{2}}M}  \label{a6}
\end{align}
where $\tau=\sin^{2}\frac{\phi}{2}$. We have calculated the three $HSSA$ up
to the next to leading order. Note that the three leading order amplitudes
in Eq.(\ref{a1}), Eq.(\ref{a3}) and Eq.(\ref{a5}) are proportional to each
other and the ratios are independent of the scattering angle (\textit{%
stringy scaling}). However, the three next to leading order amplitudes in
Eq.(\ref{a2}), Eq.(\ref{a4}) and Eq.(\ref{a6}) are NOT proportional to each
other (\textit{stringy scaling violation}).

Since $m=0$ for Eq.(\ref{a1}) and Eq.(\ref{a3}), one only needs to calculate
Eq.(\ref{00}). However since $m=1$ for Eq.(\ref{a5}), the naive order
amplitude Eq.(\ref{00}) vanishes and one needs to calculate $\frac{1}{%
\Lambda }$ order terms or Eq.(\ref{1111}) to Eq.(\ref{4444}). Similarly, to
obtain Eq.(\ref{a2}) and Eq.(\ref{a4}), one needs to calculate Eq.(\ref{1111}%
) to Eq.(\ref{4444}). To obtain Eq.(\ref{a6}), one needs to calculate $%
\frac {1}{\Lambda^{2}}$ terms in Eq.(\ref{2222}) to Eq.(\ref{3333}).

\setcounter{equation}{0} \renewcommand{\theequation}{\arabic{section}.%
\arabic{equation}}

\section{Conclusion}

Motivated by the QCD Bjorken scaling \cite{bs} and its scaling violation
correction by GLAP equation \cite{GL,AP}, in this paper, we propose a
systematic approximation scheme to calculate general string-tree level $n$%
-point $HSSA$ of open bosonic string theory. This \textit{stringy scaling
loop expansion} ($SSLE$) contains finite number of vacuum diagram terms at
each loop order of scattering energy due to a vacuum diagram contraint and a
topological graph constraint. The $4$-point leading oder results of this
calculation give the linear relations among $HSSA$ first conjectured by
Gross in 1988 \cite{GM,GM1,Gross,GrossManes} and later proved by Taiwan
group \cite{LLY2,Group,solve,LSSA} . These linear relations gave the first
evidence of the stringy scaling behavior of $HSSA$ with dim$\mathcal{M}=1$.
The $n$-point leading order results with $n\geq5$ gave the general stringy
scaling behavior of $HSSA$ with dim$\mathcal{M}=\frac{\left( r+1\right)
\left( 2n-r-6\right) }{2}$ \cite{hard,Regge}.

In addition, we give the vacuum diagram representation and its Feynman rules
for each term in the $SSLE$ of the $HSSA$. In general, there can be many
vacuum diagrams, connected and disconnected, corresponds to one term in the
expansion. Moreover, we match coefficient of each term with sum of the
inverse symmetry factors corresponding to all diagrams of the term.

Finally, as an application to extending our previous calculation of $n$%
-point leading order stringy scaling behavior of $HSSA$, we explicitly
calculate some examples of $4$-point next to leading order stringy scaling
violation terms.

At last, for future applications we summarize the power of the stringy
scaling loop expansion ($SSLE$) scheme we proposed for the calculation of $%
HSSA$ in this paper.

1. The hard scattering limit of the $4$-point function of the open bosonic
string theory can be described by vacuum diagrams of an effective field
theory with the propagator of a massless scalar field, and an infinite
number of 3, 4, 5, 6...$n$-point vertex, supplemented with the $%
u_{0}^{^{(p)}}$ factors with $u(x)$ defined in Eq.(2.14). In general, the
hard scattering limit of the n-point function ($n\geq 5$) of the open
bosonic string theory can be described by propagators of ($n-3$) massless
scaler fields, various $f_{0}^{\left( n_{2},\cdots ,n_{n-2}\right)
}=f_{0}^{\left\{ n_{i}\right\} }$ vertex in Eq.(4.12) and various $%
u_{0}^{\left( m_{2},\cdots ,m_{n-2}\right) }=u_{0}^{\left\{ m_{i}\right\} }$
factors in Eq.(4.11).

2. The $SSLE$ is in parallel to the Feynman diagram expansion for the
calculation of field theory amplitudes. However, in the $SSLE$ we give a
general formula for the coefficient of each term in the \textit{arbitrary}
higher order expansion which is difficult to calculate in the corresponding
field theory calculation. See the coefficients calculated in Eq.(\ref{sum1}%
), Eq.(\ref{term1}), Eq.(\ref{coeff}) and Eq.(\ref{000}).

3. In general, there can be many vacuum diagrams, connected and
disconnected, corresponds to \textit{one} term in the $SSLE$. This is very
different from the usual field theory Feynman expansion. Moreover, the
coefficients of these general formula for each term of $SSLE$ we calculated
is consistent with the sum of the inverse symmetry factors corresponding to
all diagrams of the term. As an example, see Eq.(\ref{000}). The calculation
of these coefficients in field theory are related to Wick theorem and
symmetry factors which are tedious to handle in the higher order field
theory expansion.

4. In contrast to the sigma-model loop expansion, or $\alpha ^{\prime }$
expansion adapted in the $\beta $ functional calculation for string in
background fields \cite{GSW} to extract \textit{low} energy effective action
of string theory, the $SSLE$ is used to extract \textit{high} energy $HSSA$
and its next to leading order or low energy corrections.

5. For a given inverse energy order $\frac{1}{\Lambda ^{M}}$ , it is
important to realize that there are only \textit{finite} number of terms in
the $SSLE$ even when dealing with massive modes due to a vacuum diagram
contraint in Eq.(\ref{66}) and a topological graph constraint in Eq.(\ref{55}%
). Moreover, we can systematically count the number of terms for each energy
order in $\frac{1}{\Lambda ^{M}}$.

However, for the usual $\alpha^{\prime}$ expansion in massive background
fields, to preserve conformal invariance in the sigma model loop
calculation, one encounters \textit{infinite} number of massive counter
terms after introducing the first massive background field. This is the
so-called non-perturbative non-renormalizability of 2-d sigma-model \cite%
{Das} and one is forced to introduce infinite number of counter-terms to
preserve the worldsheet conformal invariance \cite{Das}.

6. The $SSLE$ provides a systematic approximation scheme to calculate the
stringy scaling behavior of both $HSSA$ \cite{hard} and Regge $SSA$ ($RSSA$)
\cite{Regge} and their scaling violation terms.

7. For a given inverse energy order $\frac{1}{\Lambda ^{M}}$ , there can be $%
3,4,5,6,$...infinite many vertices in the diagramatic expansion. Moreover,
there can be many diagrams corresponding to one term in the $SSLE$. This is
much more richer than those in the usual quantum field theory expansion
which usually contains only $4$-point vertices.

In addition to the stringy scaling violation \cite{oleg}, we expect more
interesting applications of the $SSLE$ scheme to the study of high energy
string scatterings including both $HSSA$ and $RSSA$.

\section*{Acknowledgment}

We thank C. T. Chan for his early participation of some calculation of
section II. This work is supported in part by the Ministry of Science and
Technology (MoST) and S.T. Yau center of National Yang Ming Chiao Tung
University (NYCU), Taiwan.

%\bibliographystyle{unsrt}
%\bibliography{Review}

\end{document}